\documentclass[floatfix,prb,aps,superscriptaddress,10pt]{revtex4}

\usepackage{pdflscape}
\usepackage{amsfonts,amssymb}
\usepackage{amsmath}
\allowdisplaybreaks[1]
\usepackage{graphicx}
\usepackage[colorlinks,citecolor={red},linkcolor={blue}]{hyperref}
\usepackage[usenames,dvipsnames]{xcolor}
\usepackage{color}
\usepackage{ulem}
\usepackage{bm}
\usepackage{fancyvrb}

\newcommand{\refeq}[1]{Eq.~\ref{#1}}

\newcommand{\refsec}[1]{Sec.~\ref{#1}}

\newcommand{\GF}{Green's function}

\def\sxc{\Sigma_{\rm xc}}
\def\sh{\Sigma_{\rm H}}

\def\be{\begin{equation}}
\def\ee{\end{equation}}
\def\bea{\begin{eqnarray}}
\def\eea{\end{eqnarray}}

\def\y0{y_0}
\def\y00{y_0^0}
\def\etal{\textit{et al.}}

\def\beas{\begin{subequations}\begin{align}} 
\def\eeas{\end{align}\end{subequations}} 

\definecolor{MyDarkBlue}{rgb}{0,0.08,0.45}

\newcommand{\Wrpa}[1]{\mathcal{W}_{\rm RPA}[#1]}
\newcommand{\etalcomma}{\textit{et al., }}

\begin{document}

\title{Many-body perturbation theory and non-perturbative approaches: the screened interaction as key ingredient}

\author{Walter Tarantino}
 \affiliation{ Laboratoire des Solides Irradi\'es, Ecole Polytechnique, CNRS, CEA,
     Universit\'e Paris-Saclay, and 
     European Theoretical Spectroscopy Facility (ETSF), 
     F-91128 Palaiseau, France.}
\author{Bernardo S. Mendoza}
\affiliation{Centro de Investigaciones en Optica, Le\'on, Guanajuato, M\'exico}
\author{Pina Romaniello}
\affiliation{ Laboratoire de Physique Th\'eorique, IRSAMC, CNRS, 
     Universit\'e Toulouse III - Paul Sabatier and 
     European Theoretical Spectroscopy Facility (ETSF), 
     118 Route de Narbonne, F-31062 Toulouse Cedex, France}
\author{J.~A.~Berger}
\affiliation{Laboratoire de Chimie et Physique Quantiques, IRSAMC, Universit\'e Toulouse III - Paul Sabatier, CNRS and European Theoretical Spectroscopy Facility (ETSF), 118 Route de Narbonne, 31062 Toulouse Cedex, France}
\author{Lucia Reining}
 \affiliation{ Laboratoire des Solides Irradi\'es, Ecole Polytechnique, CNRS, CEA,
     Universit\'e Paris-Saclay, and 
     European Theoretical Spectroscopy Facility (ETSF), 
     F-91128 Palaiseau, France.}

\date{\today}

\begin{abstract} 
Many-body perturbation theory is often formulated in terms of an expansion in the dressed instead of the bare Green's function, and in the screened instead of the bare Coulomb interaction. However, screening can be calculated on different levels of approximation, and it is important to define what is the most appropriate choice. We explore this question by studying a zero-dimensional model (so called 'one-point model') that retains the structure of the full equations. We study both linear and non-linear response approximations to the screening. We find that an expansion in terms of the screening in the random phase approximation is the most promising way for an application in real systems. Moreover, by making use of the nonperturbative features of the Kadanoff-Baym equation for the one-body Green's function, we obtain an approximate solution in our model that is very promising, although its applicability to real systems has still to be explored.
\end{abstract}

\maketitle 










\section{Introduction}
\label{sec:intro}
In an interacting electron system, complete knowledge of the many-body wavefunction allows one
to calculate all expectation values and therefore to determine all observables.
However, in most cases it is neither possible nor desirable to calculate the full wavefunction,
as in the calculation of the observables much details of the wavefunction are integrated out.
In particular, expectation values of one-body operators and the total energy can be obtained
in a straightforward way from the one-body Green's function
$G(\mathbf{r}\sigma t,\mathbf{r}'\sigma' t')$, which 
is itself an expectation value of an electron creation and an annihilation field operator. This Green's function
can be expressed as a series
whose terms only include its non-interacting counterpart 
$G_0(\mathbf{r}\sigma t,\mathbf{r}'\sigma' t')$ 
and the Coulomb potential $v_c(\mathbf{r},\mathbf{r}')$.

Such a series, however, can be badly behaved, as, for instance, in the case
of extended systems in the thermodynamic limit.
In these situations, one can still extract sensible estimates of the Green's function
by summing up an infinite subset of terms of the series.
Cornerstone of this approach is the Dyson equation, which in a simplified notation
reads $G=G_0+G_0 \Sigma G$;
here the focus is shifted on the \emph{self-energy} $\Sigma$, whose perturbative 
expansion is still expressed in terms of $G_0$ and $v_c$ only.
Any finite truncation of the series for $\Sigma$
leads to a resummation of an infinite number of terms in the series of $G$, and only a few terms of the series of $\Sigma$ 
are usually sufficient to greatly improve on truncations of the series of $G$ itself. Still, in order to obtain satisfactory spectra, at least in an extended system it is not sufficient to keep for the self-energy only some diagrams of low order in $G_0$ and $v_c$. In particular, it has rapidly been realized that one should resum diagrams that dress the bare Coulomb interaction, since the effective interaction between charges is usually strongly screened.

In extended systems at zero temperature, the state of the art approach for the calculation of $G$, and hence of electron addition and removal spectra, is the so-called GW approximation. In a seminal paper, Hedin formalized this idea by introducing an appropriate \emph{screened interaction} $W$
and expressing the exchange-correlation contribution $\Sigma_{\rm xc}$ to the self-energy as a functional of $G$ and $W$, rather than $G_0$ and $v_c$. He derived the expression using 
Schwinger's functional derivative idea, where all the information about exchange and correlation effects on the one-body and higher order Green's functions are contained in a functional differential equation that involves the response of the system to a fictitious external potential \cite{schwinger1951a,schwinger1951b}. From the differential equation, a set of five equations are derived that are known as `Hedin's equations' \cite{hedin1965}. They contain the one-body Green's function $G$, the self-energy $\Sigma$, the irreducible polarizability $P$ which gives rise to the screened Coulomb interaction $W$, and a vertex function $\Gamma$ which stems from variations of the self-energy. 
The $GW$ approximation arises as the first step of an iterative procedure
that formally solves Hedin's equations. This first approximation corresponds to setting the vertex $\Gamma=1$. The GW approximation has encountered large success, in particular concerning the calculation of band gaps. However, in cases of failure, such as the description of Mott insulators which come out metallic in GW \cite{the_book}, or the absence of satellites related to excitations that are not due to the formation of electron-hole pairs \cite{springer1998}, a straightforward further iteration of Hedin's equations has not yet proved to be successful, and, for many applications, it is not even feasible.



The power of the GW approximation is generally attributed to the belief that the screened Coulomb interaction $W$ is weaker than the bare $v_c$, and that therefore a perturbation expansion in $W$ should be more powerful than an expansion in $v_c$. This, however, is not necessarily true. To see the point, it is sufficient to consider the screening due to a single electron in some potential. In this case the density-density response function $\chi$ is simply the non-interacting response function $\chi_0$. The inverse dielectric function $\epsilon^{-1} = 1+v_c\chi$ , which screens the Coulomb interaction via the relation $W=\epsilon^{-1}v_c$, becomes $\epsilon^{-1}=1+v_c\chi_0$. Since $\chi_0$ is negative, $W$ is indeed often smaller than $v_c$. However, for large enough (negative) $\chi_0$ compared to $1/v_c$ the inverse dielectric function and $W$ become negative, and their absolute value can be arbitrarily large. In this case the argument in favor of a perturbation expansion in terms of $W$ breaks down. Instead, this scenario never happens when $W$ is calculated in the random phase approximation (RPA), where only variations of the Hartree potential screen the interaction. In the RPA the dielectric function is $\epsilon=1-v_c\chi_0$, which is always positive and larger than 1, such that the resulting screened interaction $W_0<v_c$. This suggests that an expansion in terms of the RPA $W_0$ might be more powerful than an expansion in $W$. This indicates a first route to obtain improved expressions, which we follow in the present work.



Instead of searching for a perturbation expansion for the self-energy, one may also go back to the starting point, a functional differential equation. Here we use the Kadanoff-Baym equation (KBE) \cite{kadanoff}, which can be reformulated such that screening appears explicitly. The GW approximation is obtained from this equation as a linear-response approximation in conjunction with an approximation on variations of the Green's function with respect to the total classical potential. Also the widely used second-order cumulant approximation can be derived from this equation, again using the linear response approximation, but introducing a decoupling approximation between different states instead of the approximation on the variations of $G$. In order to go beyond both approximations, it is therefore in particular interesting to investigate contributions beyond linear response.

In the present work, both ideas are explored using a simple model, the  ``One-Point model'' (OPM).
Such a mathematical toy model, which has already been used in a similar context \cite{pavlyukh2007,molinari2005,lani2012},
represents the 0-dimensional version of a generalized Kadanoff-Baym equation.
The model captures many features of the original, multi-dimensional case;
it incorporates, for instance, the failure of the skeleton series 
that was only recently found for some multi-dimensional cases \cite{kozik,tarantino,stan}.
The model is exactly solvable, and it reflects many features of findings for real systems, such as the quality of various approximations \cite{berger2014}. Therefore, it constitutes an established first step to 
test new approximations.

The paper is organized as follows. Section \ref{sec:EOM} will be devoted to recall 
the reader the theoretical framework; our two approaches will be motivated and presented 
in Section \ref{sec:pertW0} and \ref{sec:nonepertlinear}, respectively;
results of tests on the OPM are presented in Section \ref{sec:OPM};
feasibility and computational cost of the approaches for real systems
will be discussed in Section \ref{sec:conclusions} where conclusions will also be drawn.

\section{The Equation of Motion and the Functional Approach}\label{sec:EOM}
\label{subsec:EOM}

In a non-interacting system, one can determine the one-body \GF\ $G_0$ from its
equation of motion (EOM), which is a differential equation containing the first-order time derivative of the \GF\
and the single-particle Hamiltonian. 
Together with a boundary condition in time, this fully determines  $G_0$. 
In an interacting system, the EOM for the one-body \GF\ $G$ contains additional 
contributions due to the interaction: besides the Hartree potential, a two-body \GF\ appears, 
which contains the information about exchange and correlation between the interacting particles. 
In turn, the EOM of the two-body \GF\ contains the three-body Green's function, 
and one finds an infinite chain of equations \cite{the_book}.

The equations can be cast into a more compact form 
by using the fact that higher-order \GF s can be expressed as variations of the one-body $G$ 
with respect to an external potential $\varphi$. This allows one to write the EOM for the one-body \GF\ $G$ as
 \begin{eqnarray}\label{o.15}\label{KBE} 
 G(1,2;[\varphi]) 
 &=& 
 G_0(1,2) 
 +
 \int d3 G_0(1,3) 
  \varphi(3) G(3,2;[\varphi]) 
 +
 \int d3 G_0(1,3) 
  V_\mathrm{H}(3;[\varphi]) G(3,2;[\varphi]) 
 \nonumber\\
 &+&i\int d3d4 
 \,G_0(1,3) 
  v_\mathrm{c}(3^+,4) 
 \frac{\delta G(3,2;[\varphi])}{\delta \varphi(4)}
 ,
 \end{eqnarray} 
where $v_c$ is the Coulomb potential, $G_0(1,2)=G(1,2;\varphi=0)\Big|_{v_{\rm c}= 0}$
is the Green's function of the noninteracting system, and the Hartree potential $ V_{\mathrm{H}}(1;[\varphi])$ is
 \begin{align}\label{h.1}  
 V_{\mathrm{H}}(1;[\varphi])&=
 -i\int d2 v_\mathrm{c}(1,2)G(2,2^+;[\varphi]) 
 .
 \end{align} 
 Here, we use
$(1)=(\mathbf{r}_1,\sigma_1, t_1)$
as a
short-hand notation to combine the 
space, spin, and time variables. 
Moreover, $(1^+) = (\mathbf{r}_1,\sigma_1, t^+_1 )$, 
where $t^+_1 = t_1 + \eta$ with
$\eta\to 0^+$. 

In (\ref{KBE}) the generalized one-body Green's function $G(1, 2; [\varphi])$ 
is a functional of a time-dependent external potential $\varphi(1)$. 
The equilibrium one-body Green's function of interest is retrieved in the limit of vanishing $\varphi(1)$.
Without the last term on the right-hand side of 
Eq.~\eqref{o.15}, this equation would be the Dyson equation in the Hartree approximation.  
All the many-body effects (exchange and correlation (xc)) beyond the Hartree
potential are contained in the derivative term.  

As pointed out by Baym and Kadanoff in \cite{kadanoff}, there is no known way to solve this equation. 
In particular, such a non-linear multi-dimensional functional integro-differential equation 
can have many solutions, and it may be difficult to chose which one is physical. 
Usually, the problem is circumvented by using some kind of iterative approach,
starting from the non-interacting or the Hartree solution.
A straightforward iteration of Eq.~\eqref{o.15} leads to series of terms of increasing order 
in the interaction and corresponds to a perturbation expansion of $G$. 
However, such a series has often bad convergence properties 
(see also \refsec{sec:OPM} for an illustration).
Therefore a common way to go is to transform Eq.~\eqref{o.15} into a Dyson equation:
\begin{equation}\label{eq:dyson} 
 G(1,2;[\varphi]) 
 = 
 G_H(1,2;[\varphi])  
 +i
 \,G_H(1,\bar 3;[\varphi]) \Sigma_{\rm xc}(\bar 3,\bar 6;[\varphi])
 G(\bar 6,2;[\varphi])
 ,
\end{equation}
with the exchange-correlation self-energy
\begin{equation}
 \Sigma_{\rm xc}(3,6;[\varphi]) \equiv  -i v_\mathrm{c}(3^+,\bar 4) G(3,\bar 5;[\varphi])\frac{\delta G^{-1}(\bar 5, 6;[\varphi])}{\delta \varphi(\bar 4)}.
 \label{eq:sigma}
\end{equation}
Here and in the following, we denote integrals by $f(\bar n)g(\bar n)\equiv \int dn\,f(n)g(n)$.


Once the expression for the self-energy is established,  
it is enough to solve \refeq{eq:dyson}  for $\varphi \to 0$ 
in order to obtain the equilibrium \GF . 
Of course, since $G$ is unknown so is $\Sigma_{\rm xc}$: 
with the transformation of the EOM into a Dyson equation, the goal becomes to find accurate
approximations to the self-energy. 
The advantage of the Dyson equation is the fact that even a low-order approximation
to the self-energy creates in the \GF\ terms to infinite order in the interaction, 
so one may hope that approximating $\Sigma_{\rm xc}$ is easier than approximating $G$ itself. 

One strategy to find approximations for $\Sigma_{\rm xc}$ has been formalized by Hedin \cite{hedinPR65}, based on the insight that, at least for extended systems, \textit{screening} of the interaction plays a key role. The screening is due to the self-consistent reaction of the electron system to the variation of the external potential. This self-consistency is automatically taken into account when the variation of the \textit{total}, i.e. external plus system-internal, potential is considered. Hedin's formulation uses the classical part of the total potential, $\varphi_{\rm cl}=\varphi + V_H$, to express the exact exchange-correlation self-energy (\ref{eq:sigma}) as
\begin{equation}
 \Sigma_{\rm xc}(3,6;[\varphi]) \equiv  -i v_\mathrm{c}(3^+,\bar 4) G(3,\bar 5;[\varphi])\frac{\delta G^{-1}(\bar 5, 6;[\varphi_{\rm cl}])}{\delta \varphi_{\rm cl}(\bar 7)}\frac{\delta \varphi_{\rm cl}(\bar 7)}{\delta \varphi(\bar 4)}\equiv iG(3\bar 4)W(3^+,\bar 5)\tilde\Gamma(\bar 4,6;\bar 5)
 \label{eq:sigma-2}
\end{equation}
with
\begin{equation}
\label{eq:WDOP}
W(1,2) = v_c(1,2) + v_c(1,\bar 3)P(\bar 3,\bar 4)W(\bar 4,2)
\end{equation}
and the irreducible polarizability
\begin{equation}
\label{eq:PG}
P(1,2) = -i   G(1,\bar 3)G(\bar 4,1)\tilde\Gamma(\bar 3,\bar 4;2).
\end{equation}
From the Dyson equation (\ref{eq:dyson}), the fifth equation in Hedin's set is derived for the irreducible vertex function $\tilde \Gamma \equiv -\delta G^{-1}/\delta \varphi_{\rm cl}$ as
\begin{equation}
\label{eq:GAMMA}
\tilde\Gamma(1,2;3) = \delta(1,2)\delta(1,3)+
  \frac{\delta
 \sxc(1,2)}{\delta G(\bar 4,\bar 5)}
 G(\bar 4,\bar 6)G(\bar 7,\bar 5)\tilde\Gamma(\bar 6,\bar 7;3).
 \end{equation}
The screened Coulomb interaction $W$ is linked to the density-density response function
\begin{equation}
\chi(1,2) \equiv \frac{\delta n(1)}{\delta \varphi(2)}
\end{equation}
via $W(1,2) = v_c(1,2) + v_c(1,\bar 3) \chi(\bar 3,\bar 4) v_c(\bar 4,2)$,
and it can also be expressed in terms of the inverse dielectric function $\epsilon^{-1}(1,2)\equiv \delta(1,2)+v_c(1,\bar 3)\chi(\bar 3,2)$ as
\begin{equation}
W(1,2) = \epsilon^{-1}(1,\bar 3)v_c(\bar 3,2).
\end{equation}

By setting $\tilde \Gamma\approx 1$, the equations reduce to a closed set of algebraic equations
for $G$, $\Sigma$, $W$ and $P$, leading to the so called GW approximation (GWA) for the self-energy and the Random Phase Approximation (RPA) for $P$.
The equations can now be solved self-consistently. Often, however, some mean-field \GF\ is used instead of the self-consistent $G$ to build the self-energy. The mean-field \GF\ is usually referred to as $G_0$, although it contains some interaction through, e.g., the KS potential or the Fock exchange. This approach is usually called $G_0W_0$.
For later reference we highlight the fact that $W$ is calculated in the RPA by using the notation 

\begin{equation}\label{oneshotGW}
 \Sigma_{\rm xc}\approx iG_0\Wrpa{G_0} \,\,\,\,\,\,\,\,\,\,\,\,\,\,\,
  W\approx \Wrpa{G_0}\ \,\,\,\,\,\,\,\,\,\,\,\,\,\,\,
  P \approx -iG_0G_0 \,\,\,\,\,\,\,\,\,\,\,\,\,\,\,
  \tilde \Gamma \approx 1 ,
\end{equation}
where the \textit{RPA functional} is $\Wrpa{F}\equiv (1+iFFv_c)^{-1}v_c$.
Such a functional generalizes the functional form of $W$ obtained 
from the Random Phase Approximation for $P$, where $\tilde\Gamma =1$ implies $P\approx-iGG$
and hence $W\approx (1+iv_c GG)^{-1}v_c$.

The $G_0W_0$ scheme has led to good results in many cases, but it suffers from the fact that these results strongly depend on the choice of the mean field in $G_0$. For this reason, partially self-consistent schemes have been developed and led to important progress. Technically, even fully self-consistent GW calculations are nowadays feasible \cite{caruso2012}. However, it is not clear whether full self-consistency on the GW level is desirable; in particular, it has been pointed out that the resulting $\chi$  does not fulfill the $f$-sum rule \cite{holm1998}. Clearly, vertex corrections, i.e. corrections to the self-energy of higher order in $W$, are needed to improve this situation. 

\section{Perturbation theory in  \texorpdfstring{$W_0$}{W0}}\label{sec:pertW0}
The success of the GW approximation may be explained by
the intuitive picture according to which the screened potential $W$
represents a better parameter than the bare Coulomb interaction $v_c$
for a perturbative expansion of $\Sigma_{xc}$. This presumes that 
$|W|< v_c$. Although often true, this is simplistic, and as explained in the introduction $|W|$ can in principle become arbitrarily large. 
On the other hand, $P^{RPA}(\omega=0;[G_0])$ built with any mean-field $G_0$ is always negative, which means that  $\Wrpa{G_0}$ is always positive and smaller than $v_c$. Moreover, the corresponding density-density response function $\chi^{RPA}[G_0]$ is well behaved; in particular, it fulfills the $f$-sum rule.


Based on these considerations, we propose to consider the perturbative expansion
of $\Sigma_{xc}$ in $G_0$ and $\Wrpa{G_0}$. 
This is formally constructed in the following way.
First, we use the definition of $\Wrpa{G_0}$ to invert the map $v_c \to \Wrpa{G_0}$, which leads to 
$v_c=(\Wrpa{G_0}^{-1} + P_0)^{-1}= \Wrpa{G_0}(1-i\Wrpa{G_0}G_0G_0)^{-1}$
with $P_0 \equiv -iG_0G_0$.
We then use the well established perturbation expansion of $\Sigma_{xc}$ in 
terms of $v_c$ and insert the above expression for $v_c$ to
rewrite the series in terms of $\Wrpa{G_0}$ and, finally,
we group terms of the same order in $\Wrpa{G_0}$. By defining $W_0\equiv \Wrpa{G_0}$
we can schematically represent the procedure as
\begin{multline}
\Sigma_{xc}=iv_c G_0+v_c G_0 v_c G_0G_0+...\\
    = i (W_0^{-1} + P_0)^{-1}G_0 
    +(W_0^{-1} + P_0)^{-1}G_0(W_0^{-1} + P_0)^{-1}
        	G_0 G_0+...\\
   	=iW_0G_0+
      W_0( G_0 W_0 G_0G_0
      -G_0G_0 W_0G_0)+...
\end{multline}
The first order yields $G_0W_0$, while the second order modifies the Second Born approximation.
More in detail, the first order contribution to the self-energy is
\be
\sxc^{(1)}(1,2) =i W_0(1,2)G_0(1,2),
\ee
and the second order contributions read
\be 
\sxc^{(2)}(1,2) =\Sigma_2(1,2;[W_0])
-W_0(1,\bar{3})G_0(\bar{3},\bar{4})G_0(\bar{4},\bar{3})W_0(\bar{4},2)G_0(1,2),
\ee
where $W_0(1,2)\equiv \left(1-v_c\chi_0\right)^{-1}(1,\bar 3)v_c(\bar 3,2)$
and $\Sigma_2(1,2;[W_0])$ are all second order diagrams of the expansion in $v_c$ evaluated with $W_0$
instead of $v_c$. Note the subtraction of the last diagram, which is necessary in order to avoid double counting of the bubble diagrams.

\section{Nonperturbative Linear Response}\label{sec:nonepertlinear}

The previous section designs a way to use a linear response ingredient, namely $W_0$, in order to build more powerful perturbation series. In the present section we will discuss a way to make non-linear response appear.
Since the expressions become quickly very clumsy, we suppress in the presentation all arguments except for the functional dependence on the fictitious external potential, which is crucial for the derivations.


Starting from (\ref{o.15}), we focus on the self-energy
\be\label{sigma.ini}
\Sigma[\varphi] =\varphi + V_H[\varphi]+iv_c\frac{\delta G[\varphi]}{\delta \varphi}G^{-1}[\varphi].
\ee
We assume that $V_H[\varphi]$ (defined in (\ref{h.1})) is Taylor expandable:
\be\label{VTaylor}
V_H[\varphi]=-iv_c n-i \varphi v_c \frac{\delta G[\varphi]}{\delta \varphi}\Big|_{\varphi\to0}
-\frac{i}{2}\varphi\varphi v_c \frac{\delta^2 G[\varphi]}{\delta \varphi\delta \varphi}\Big|_{\varphi\to0}+...
\ee
where $n(1)\equiv G(1,1^+;[0])$ is the charge density.
As first approximation, we truncate the above series to first order
\be\label{approx1}
V_H[\varphi]\approx -iv_c n-i \varphi v_c \frac{\delta G[\varphi]}{\delta \varphi}\Big|_{\varphi\to0}
\ee
which leads to
\be\label{sigma.1app}
\Sigma[\varphi] \approx -iv_c n+\varphi \left(\delta-i v_c \frac{\delta G[\varphi]}{\delta \varphi}\Big|_{\varphi\to0}\right)
										+iv_c\frac{\delta G[\varphi]}{\delta \varphi}G^{-1}[\varphi].
\ee
This rewriting suggests to renormalize (or \textit{screen}) the external potential as
\be\label{barphi}
\varphi\to \bar{\varphi}\equiv \varphi \left(\delta-i v_c \frac{\delta G[\varphi]}{\delta \varphi}|_{\varphi\to0}\right).
\ee
Since (\ref{barphi}) implies 
\bea
\frac{\delta G[\varphi]}{\delta \varphi}
	&=&\frac{\delta \bar{G}[\bar{\varphi}]}{\delta \bar{\varphi}}
     \left(\delta-iv_c\frac{\delta G[\varphi]}{\delta \varphi}\Big|_{\varphi\to 0} \right)\\
\Rightarrow \frac{\delta G[\varphi]}{\delta \varphi}\Big|_{\varphi\to 0}
	&=&\frac{\delta \bar{G}[\bar{\varphi}]}{\delta \bar{\varphi}}\Big|_{\bar{\varphi}\to 0 \Leftrightarrow \varphi\to 0}
     \left(\delta-iv_c\frac{\delta G[\varphi]}{\delta \varphi}\Big|_{\varphi\to 0} \right)     \\
\Rightarrow \frac{\delta G[\varphi]}{\delta \varphi}\Big|_{\varphi\to 0}
	&=&
     \left(\delta+iv_c\frac{\delta \bar{G}[\bar{\varphi}]}{\delta \bar{\varphi}}\Big|_{\bar{\varphi}\to 0} \right)^{-1}
       \frac{\delta \bar{G}[\bar{\varphi}]}{\delta \bar{\varphi}}\Big|_{\bar{\varphi}\to 0}\\
\Rightarrow      \frac{\delta G[\varphi]}{\delta \varphi}
	&=& \frac{\delta \bar{G}[\bar{\varphi}]}{\delta \bar{\varphi}}
    	\left(\delta-iv_c
          \left(
        	\delta+i v_c  \frac{\delta \bar{G}[\bar{\varphi}]}{\delta \bar{\varphi}}\Big|_{\bar{\varphi}\to 0}
            	\right)^{-1} 
                 \frac{\delta \bar{G}[\bar{\varphi}]}{\delta \bar{\varphi}}\Big|_{\bar{\varphi}\to 0} 
                   \right)
\eea
with $\bar{G}[\bar{\varphi}]\equiv G[\varphi[\bar{\varphi}]]$,
we can write (\ref{sigma.1app}) as
\be
\bar{\Sigma}[\bar{\varphi}]\equiv \Sigma[\varphi[\bar{\varphi}]] \approx -iv_c n+\bar{\varphi}+iv_c \frac{\delta \bar{G}[\bar{\varphi}]}{\delta \bar{\varphi}}
    	\left(\delta-iv_c
          \left(
        	\delta+i v_c  \frac{\delta \bar{G}[\bar{\varphi}]}{\delta \bar{\varphi}}\Big|_{\bar{\varphi}\to 0}
            	\right)^{-1} 
                 \frac{\delta \bar{G}[\bar{\varphi}]}{\delta \bar{\varphi}}\Big|_{\bar{\varphi}\to 0} 
                   \right)G^{-1}[\varphi].
\ee
If one now approximates the functional derivative as
\be\label{approx2}
\frac{\delta \bar{G}[\bar{\varphi}]}{\delta \bar{\varphi}}\approx \bar{G}[\bar{\varphi}]\bar{G}[\bar{\varphi}]
\ee
one obtains 
\be
\bar{\Sigma}[\bar{\varphi}] \approx -iv_c n+\bar{\varphi}+iv_c \bar{G}[\bar{\varphi}]\bar{G}[\bar{\varphi}]
    	\left(\delta-iv_c
          \left(
        	\delta+i v_c  \bar{G}[0]\bar{G}[0]
            	\right)^{-1} 
                 \bar{G}[0]\bar{G}[0]
                   \right)G^{-1}[\varphi].
\ee
Since the derivative part has been approximated, this equation is 
no longer a functional-differential equation but a Dyson equation,
which can be solved directly at $\varphi=\bar{\varphi}=0$.
The equilibrium self-energy is then obtained as
\be\label{sigma.almostrpa}
\bar{\Sigma}[0]=\Sigma[0]\approx V_H[0]
			+iv_c \left( 1
            -i G[0] G[0]v_c
          \left(
        	1+i v_c  G[0]G[0]
            	\right)^{-1} \right)
                 G[0]
\ee
since $\bar{G}[0]=G[0]$.  
Here the RPA screened interaction appears, since 
\be
\mathcal{W}_{\rm RPA}[X]=v_c \epsilon^{-1}_{\rm RPA}[X]=v_c(1+v_c \chi_{\rm RPA}[X] )		\equiv v_c\left(\delta + v_c \left(-iXX(\delta+iXX v_c)^{-1}\right)\right).
\ee
Since $\bar{G}[0]=G[0]$ the self-energy (\ref{sigma.almostrpa}) becomes
\be\label{sigma.AD}
\Sigma[0] \approx V_H[0]+i\mathcal{W}_{\rm RPA}[G[0]] G[0]
	\approx V_H +i\mathcal{W}_{\rm RPA}[G_0] G_0.
\ee
These are, respectively, the self-consistent $GW$ approximation with $W\approx\Wrpa{G}$, and the so-called $G_0W_0$ approximation, where the expressions are evaluated with non-interacting Green's functions; in practice, most often some mean-field solution such as a Kohn-Sham Green's function is used for $G_0$.


The above derivation of the $G_0W_0$ approximation relies 
on two ordered approximations: (\ref{approx1}) and (\ref{approx2}),
that are tied to each other via the definition of the screened
perturbing potential $\bar{\varphi}$. 
The first one makes the problem linear, while the second one
removes the differential character and leads to an algebraic equation.
If one would apply (\ref{approx2}) directly on (\ref{sigma.ini}),
without going through (\ref{approx1}) and the definition of $\bar{\varphi}$,
one would get to
\be
\Sigma[\varphi]\approx \varphi + V_H[\varphi]+iv_c G[\varphi]
\ee
which is simply the Hartree-Fock approximation.\\
The derivation seems to suggest that a better approximation may come from improving on (\ref{approx1}),
namely taking more orders of the Taylor expansion of $V_H[\varphi]$.
However, solving the Dyson equation at equilibrium kills all possible improvements.
Such a mechanism is clear already when keeping the second order in (\ref{VTaylor}).
Let us first write
\be
\Sigma[\varphi]\approx -iv_c n+\varphi \left(\delta-i v_c \frac{\delta G[\varphi]}{\delta \varphi}\Big|_{\varphi\to0}
				-\frac{i}{2}\varphi v_c \frac{\delta^2 G[\varphi]}{\delta \varphi\delta \varphi}\Big|_{\varphi\to0}\right)
										+iv_c\frac{\delta G[\varphi]}{\delta \varphi}G^{-1}[\varphi].
\ee
Now we introduce the screened perturbing potential (in the even shorter notation for which functional derivatives $\delta F[\varphi]/\delta \varphi$ are denoted by $F'[\varphi]$)
\be
\tilde{\varphi}\equiv \varphi(1-iv_cG'[0]-\frac{i}{2}v_c\varphi G''[0])
\ee
for which
\be
G'[\varphi]=(1-iv_cG'[0]-\frac{i}{2}\varphi v_c G''[0]-\frac{i}{2}\varphi v_c G''[0])\tilde{G}'[\tilde{\varphi}]
\ee
where $\tilde{G}[\tilde{\varphi}]\equiv G[\varphi[\tilde{\varphi}]]$.
This implies that
\be
G'[0]=\tilde{G}[0](1+iv_c\tilde{G}'[0])^{-1}
\ee
which, in combination with 
\be\label{approx2.1}
\frac{\delta \tilde{G}[\tilde{\varphi}]}{\delta \tilde{\varphi}}\approx \tilde{G}[\tilde{\varphi}]\tilde{G}[\tilde{\varphi}],
\ee
gives
\be
\bar{\Sigma}[\bar{\varphi}]\equiv \Sigma[\varphi[\bar{\varphi}]]\approx -iv_c n+\tilde{\varphi}+iv_c\left(1
	-i\tilde{G}[0]\tilde{G}[0](1+iv_c \tilde{G}[0]\tilde{G}[0])^{-1}v_c\right)
		\tilde{G}[\tilde{\varphi}]
\ee
which at equilibrium is just
\be
\bar{\Sigma}[0]=\Sigma[0]\approx -i v_c \tilde{G}[0]+i\mathcal{W}_{RPA}[\tilde{G}[0]] \tilde{G}[0].
\ee
It follows that the only way to include non-linear screening effects is to
go beyond the second approximation (\ref{approx2},\ref{approx2.1}).
This suggests that in order to go beyond the GWA one should keep the functional derivative, and attempt to solve the
linearized functional-differential equation itself, as it has been suggested in \cite{lani2012} and pursued in \cite{guzzo2011,sky2017} in the context of cumulant approximations.

According to the order of the truncation of \eqref{VTaylor}, we get different
approximate KBEs, which we shall collectively refer to as `hierarchy of linearized KBEs', since these differential equations are linear in the variable $G[\varphi]$.
The coefficients of the Taylor expansion can in principle be estimated within Time-dependent Density Functional Theory,
or evaluated self-consistently once the differential equation is solved.\footnote{%
The entire procedure will be made more clear in the next section,
where it will be implemented on a toy model.}
The hope is that one recovers the original KBE by climbing the hierarchy of linearized KBEs,
namely by considering more and more terms of \eqref{VTaylor}, and that the resulting hierarchy of physical solutions to these equations smoothly converges to the exact Green's function.

\section{Tests on the OPM}\label{sec:OPM}
Solving the functional-differential
equation \eqref{o.15}, even with the simplest approximations for $V_H[\varphi]$,
is a tough problem. First attempts in this direction were carried out by Lani 
and collaborators \cite{lani2012}, who found it fruitful to study the equation in its
zero-dimensional version, in which case the equation becomes a simple
differential equation for a scalar function of one variable. 
Later on, Berger and collaborators \cite{berger2014} generalized such a differential
equation in a way to describe more physics by keeping the same simple framework.
The resulting equation, usually referred to as the `one-point model' (OPM),
encodes highly nontrivial features of the functionals discussed in Section \ref{subsec:EOM},
which otherwise could be studied only in their perturbative regime and 
only for a very limited order of expansion, due to the
factorially increasing number of terms produced by perturbation theory.
Despite its simplicity, the OPM has proved to be a surprisingly reliable benchmark  
for qualitative analysis of approximation schemes, as proved by the analyses performed by
Berger \etal\ and, later, Stan \etalcomma who also showed that the model can be used to
study pathologies of Perturbation Theory such as the 
misconvergence of the skeleton series \cite{stan,kozik}.
This makes the model particularly appropriate for the study
of the nonperturbative features of the equation of motion and its approximations, discussed in the previous section. Moreover, one can also use it to test perturbative schemes like the one based on an expansion in the RPA screened interaction,
with the possibility of exploring at glance a large number of terms of a series
and getting qualitative insights concerning its convergence.

In the next subsections, we shall recall the relevant ingredients of the OPM 
(\ref{subsec:OPM});
show the performance of the OPM equivalent of GWA (\ref{subsec:GWA}), 
which will serve as benchmark for our suggested approximations;
derive and study the series in the equivalent of $\Wrpa{G_0}$
and compare it with the series in the exact $W$ (\ref{subsec:wrpa});
solve the hierarchy of the OPM equivalent of the linearized KBE equations (\ref{subsec:nplr});
and, finally, compare all approaches (\ref{subsec:comp}).

\subsection{The One-Point Model}\label{subsec:OPM}

We here summarize the relevant features of the OPM. The reader interested 
to more details is referred to the work of Berger \etal \cite{berger2014}.
At the basis of the model we find the differential equation:
\be
y(x)=y_0(x)+y_0(x)\left(-v  y(x)+\lambda  v  y'(x)y(x)^{-1}\right)y(x)
\ee
with $y_0(x)\equiv y_0^0/(1-y_0^0 x)$, $x\in \mathbb{R}$, $\lambda>0$, $v\ge 0$, $\y00>0$.
For $\lambda=1$, the above equation corresponds to the zero-dimensional
version of the equation of motion \eqref{KBE}, provided that
\begin{subequations}\label{subs1}
\begin{align}
G(1,2;[\varphi])&\to iy(x)
, 
\\
G_0(1,2)&\to iy_0^0
, 
\\
\varphi(1)&\to -ix
,
\\
v_c&\to iv 
\end{align}
\end{subequations}
For this specific value of $\lambda$, the exact problem is characterized by the trivial
physical solution $y(x)=y_0(x)$.
A much reacher result is obtained by considering $\lambda \neq 1$. Here we take $\lambda=1/2$, namely
\be\label{opmkbe}
y(x)=y_0(x)+y_0(x)\left(-v  y(x)+\frac{1}{2} v  y'(x)y(x)^{-1}\right)y(x).
\ee
The choice $\lambda \neq 1$ reflects the fact that in the real system Hartee and exchange contributions do not cancel perfectly, whereas this would be the case in the model for $\lambda =1$. In this sense, $\lambda=1/2$ is linked to the prefactor $1/2$ that is often used in real systems for exchange terms when spin is not explicitly considered.

\begin{figure}[h]
\includegraphics[scale=0.7]{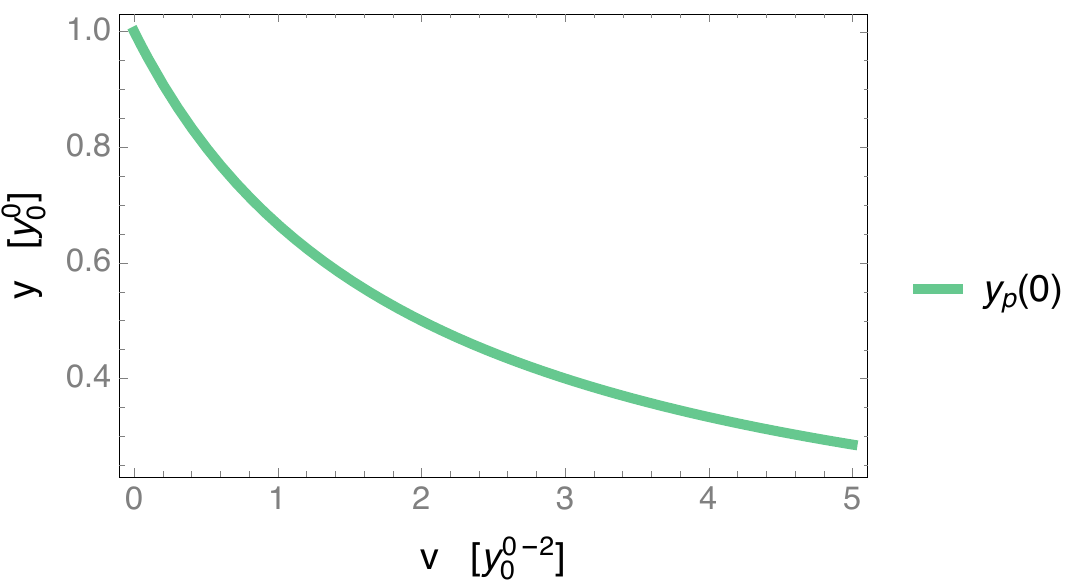}
\caption{The function $y_p(x)$ represents the functional $G[\varphi]$, therefore the 
equilibrium Green's function, which is the values of the functional at $\varphi=0$,
is represented by $y_p(0)$. A specific value of $y_p(0)$ is determined by $v$, 
which represents the coupling between the electrons, and $y_0^0$, which represents
the non-interacting Green's function and encodes all information of the non-interacting system. 
In analogy with the study of the Hubbard model, we can consider fixed 
(say $y_0^0=1$) and let $v>0$.
While varying, $v$ spans a `weakly coupled regime' $v\approx 0$, a `strongly coupled regime' $v\gg 0$, and an intermediate region.
The notation `$y\;[y_0^0]$' and `$v\;[y_0^0{}^{-2}$', here and in all following plots,
is a reminder that the problem can be entirely rewritten in terms of the adimensional quantities $y/y_0^0$ and $v y_0^0{}^2$
and therefore $y$ can be expressed in units of $y_0^0$ and $v$ in units of $y_0^0{}^{-2}$.
}
\label{fig:OPM}
\end{figure}

In analogy with the original problem, one can use the equation to get the perturbative
expansion of $y(x)$ in $v$ without actually solving the differential problem.
In fact, the resulting series:
\be\label{opmpt}
y(x)=y_0(x)-\frac{1}{2} v y_0(x){}^3+\frac{1}{4} v^2 y_0(x){}^5-\frac{1}{8} v^3 y_0(x){}^7+\frac{1}{16} v^4 y_0(x){}^9+...
\ee
can be summed up to the function
\be\label{opmphys}
y_p(x)\equiv \frac{y_0(x)}{\frac{1}{2} v y_0(x){}^2+1}
\ee
which is the analytic continuation of the series \eqref{opmpt}.
Since \eqref{opmphys} is also solution to \eqref{opmkbe},
we identify this as the \emph{physical solution} of the problem,
all other solutions to \eqref{opmkbe} being labeled as 
unphysical ones. 
Such a function is a model of the functional that in the
standard many-body problem realizes the map $\varphi \to G$ by analytic continuation
of the series produced by Perturbation Theory.
The Green's function at equilibrium, namely $G[\varphi = 0]$, is then represented by
\be\label{exacty}
y_p(0)=\frac{y_0^0}{\frac{y_0^0{}^2 v}{2}+1},
\ee
which is plotted in fig. \ref{fig:OPM}.
The function $y_0(x)$, to which $y_p(x)$ reduces for $v\to 0$, is then identified as
the out-of-equilibrium non-interacting Green's function, the corresponding version at equilibrium 
being simply $y_0(0)=\y00$.
Provided with these functions, we can define the \emph{physical self-energy} $\Sigma$,
the \emph{Hartree self-energy} $\sh$, the \emph{exchange-correlation self-energy}
$\sxc$, the \emph{screened potential} $w$ and the \emph{RPA function(al)} $w_{rpa}(f)$
as
\begin{subequations}
\begin{align}
\Sigma&\equiv y_0(0)^{-1}-y_p(0)^{-1}=-\frac{y_0^0 v}{2}\\
\sh&\equiv-vy_p(0)=-\frac{y_0^0 v}{\frac{y_0^0{}^2 v}{2}+1}\\
\sxc&\equiv \Sigma -\sh =\frac{y_0^0 v \left(2-y_0^0{}^2 v\right)}{2 y_0^0{}^2 v+4}\\
p&\equiv y_p'(0)\\
\epsilon^{-1}&\equiv 1-v p \\
w&\equiv \epsilon^{-1}v=\frac{v \left(3 y_0^0{}^4 v^2+4\right)}{\left(y_0^0{}^2 v+2\right){}^2}\label{Wexact}\\
W_{rpa}(f)&\equiv\frac{v}{f^2 v+1}.\label{wrpadef}
\end{align}
\end{subequations}
In the rest of the section we shall consider $\Sigma_H$ as given once for all.
This is different from the work of Berger \etal\  
where $\Sigma_H$ was calculated self-consistently. We make this choice since it 
will allow us to better represent standard many-body techniques, and $G_0 W_0$ in
particular, in which $\Sigma_H$ is usually a given input coming
from a DFT calculation. Moreover it allows us to discuss the two proposed schemes in a consistent way.

\subsection{The GWA}\label{subsec:GWA}

To assess the quality of the proposed approximations, we shall compare them with the OPM
equivalent of the GWA. A thorough analysis of such an approximation was worked out by 
Berger \etal \cite{berger2014}, who considered a self-consistently calculated $\sh$. As pointed out above, here 
we make an analogous analysis, but with a $\sh$ which is supposed to be given.
The self-consistent $GW$ approximation corresponds to solving the set of equations
\begin{subequations}
\begin{align}
P & \approx  y^2 \\
W & \approx  v-vPW \\
y&\approx \y00+\y00(\sh+\frac{1}{2}y W)y\label{scGWdyson}.
\end{align}
\end{subequations}
This set of equations admits 3 solutions, of which only the following reduces to $\y00$
in the noninteracting limit:
\be\label{ygw}
y_{gw}=\frac{\frac{3\ 3^{2/3} v \left(3 y_0^0{}^2 v \left(y_0^0{}^2 v \left(y_0^0{}^2 v-8\right)-12\right)-16\right)}{\sqrt[3]{B}}+3 \sqrt[3]{3} \sqrt[3]{B}+9 y_0^0 v \left(y_0^0{}^2 v+2\right)}{18 v \left(3 y_0^0{}^2 v+2\right)}
\ee
with 
\bea
A&\equiv& \sqrt{3 y_0^0{}^2 v \left(3 y_0^0{}^2 v \left(y_0^0{}^2 v \left(3 y_0^0{}^2 v \left(y_0^0{}^2 v \left(2 y_0^0{}^2 v+13\right)+80\right)+488\right)+416\right)+464\right)+256}\\
B&\equiv& 2 \sqrt{3} A v^{3/2} \left(3 y_0^0{}^2 v+2\right)+9 y_0^0{}^9 v^6+216 y_0^0{}^7 v^5+648 y_0^0{}^5 v^4+576 y_0^0{}^3 v^3+144 y_0^0 v^2.
\eea

For real systems a direct solution of the $GW$ equations is, in general, not feasible. Therefore, in practice, they are usually solved iteratively. Since the $GW$ equations can be rewritten in several ways, various iterative schemes are possible~\cite{berger2014}.
Interestingly, the above solution can be obtained by the following iterative procedure:
\begin{subequations}
\begin{align}
P_i & =  y_i^2 \\
W_i & =  v-vP_iW_i \\
y_{i+1}&=\y00+\y00(\sh+\frac{1}{2}y_i W_i)y_{i+1}
\end{align}
\end{subequations}
starting from $y_1=\y00$,
which is the standard iterative scheme used in $GW$ calculations for real systems. We note that the above iterative scheme converges to the physical solution for all $v$ thanks to the fact that we consider $\Sigma_H$ fixed. Otherwise the above iterative scheme does not converge for large $v$~\cite{berger2014}.

Since we can also calculate the exact screened potential $W$,
it is worthwhile to explore also a $GW$ self-energy built with the exact $W$,
which could in principle be obtained from TDDFT.
In this case, we solve the Dyson equation Eqn.\eqref{scGWdyson} 
with the exact screened potential $W=w$ of \eqref{Wexact}. Of the two solutions,
the only one that goes to zero in the limit $v \to 0$ is
\be
y_{gwexact}=\frac{\left(y_0^0{}^2 v+2\right) \left(3 y_0^0{}^2 v-\sqrt{-6 y_0^0{}^6 v^3+9 y_0^0{}^4 v^2+4 y_0^0{}^2 v+4}+2\right)}{y_0^0 v \left(3 y_0^0{}^4
   v^2+4\right)}.
\ee

\begin{figure}[h]
\includegraphics[scale=0.7]{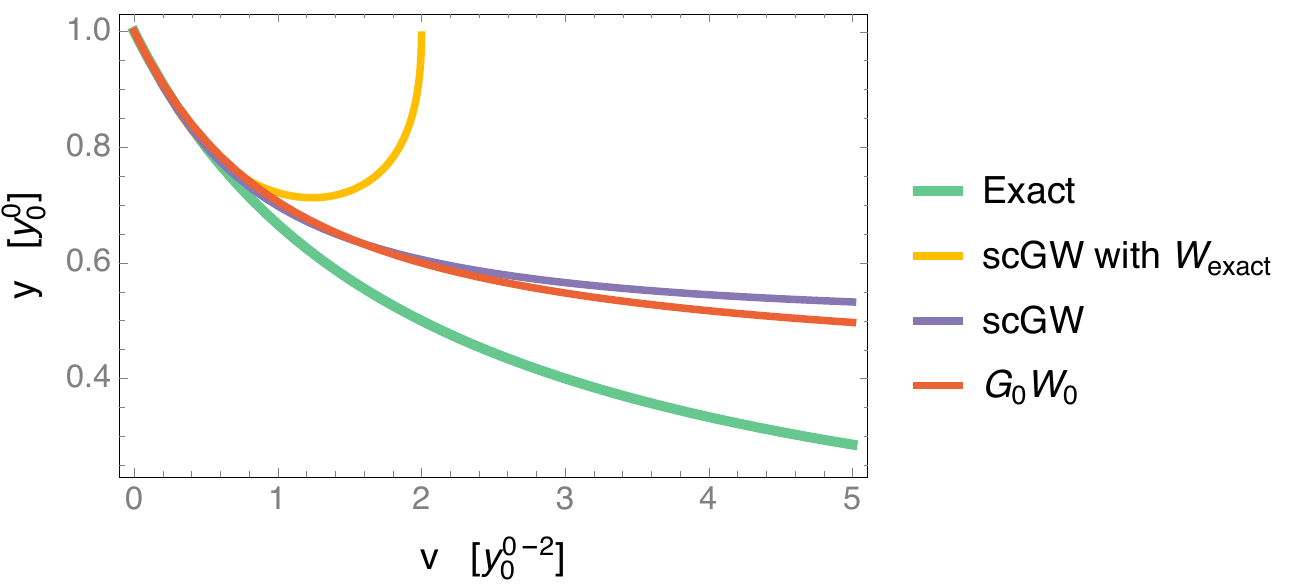}
\caption{Comparison of the Green's function calculated using various flavours of the $GW$ approximation with the exact solution, as a function of the interaction.}
\label{fig:gwcompare}
\end{figure}

In figure \ref{fig:gwcompare}, the OPM equivalent of the self-consistent $GW$
approximation and self-consistent $GW$ with $W$ exact are shown in comparison with
the equivalent of $G_0W_0$. In analogy with what is found in realistic calculations, self-consistent $GW$ does not
improve systematically on $G_0W_0$ and is clearly worse for larger values of the interaction.

\subsection{Perturbation Theory in \texorpdfstring{$W_0$}{W0} and \texorpdfstring{$W$}{W}}\label{subsec:wrpa}

\begin{figure}[h]
\includegraphics[scale=0.4]{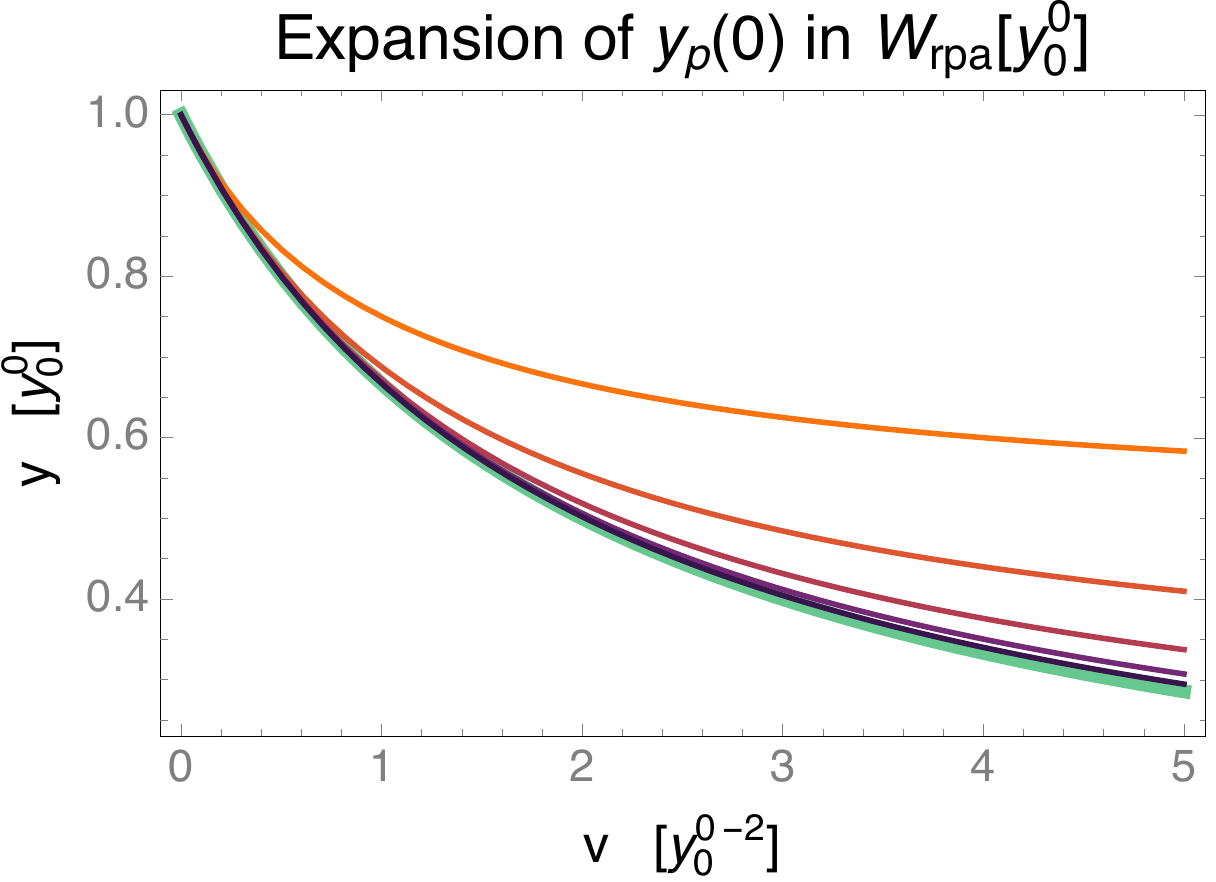}
\includegraphics[scale=0.4]{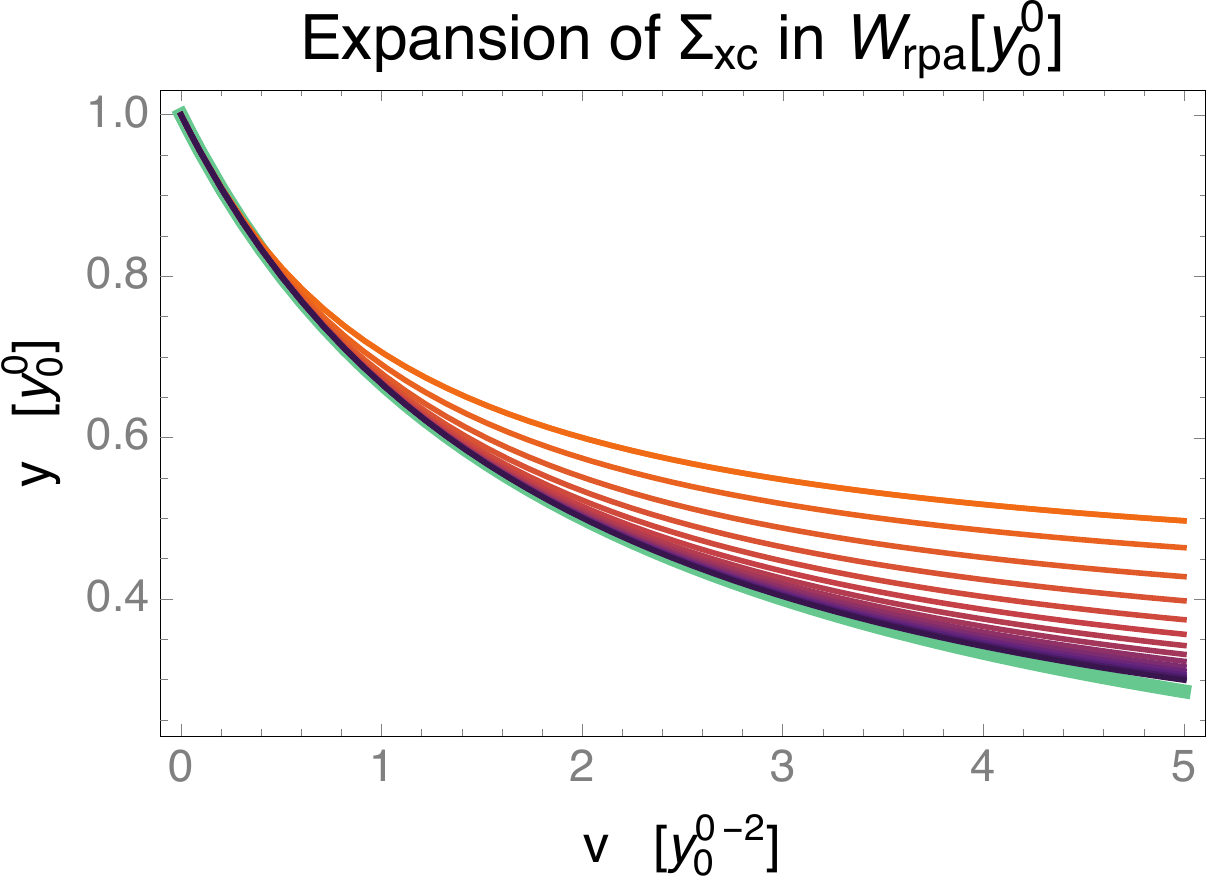}
\includegraphics[scale=0.4]{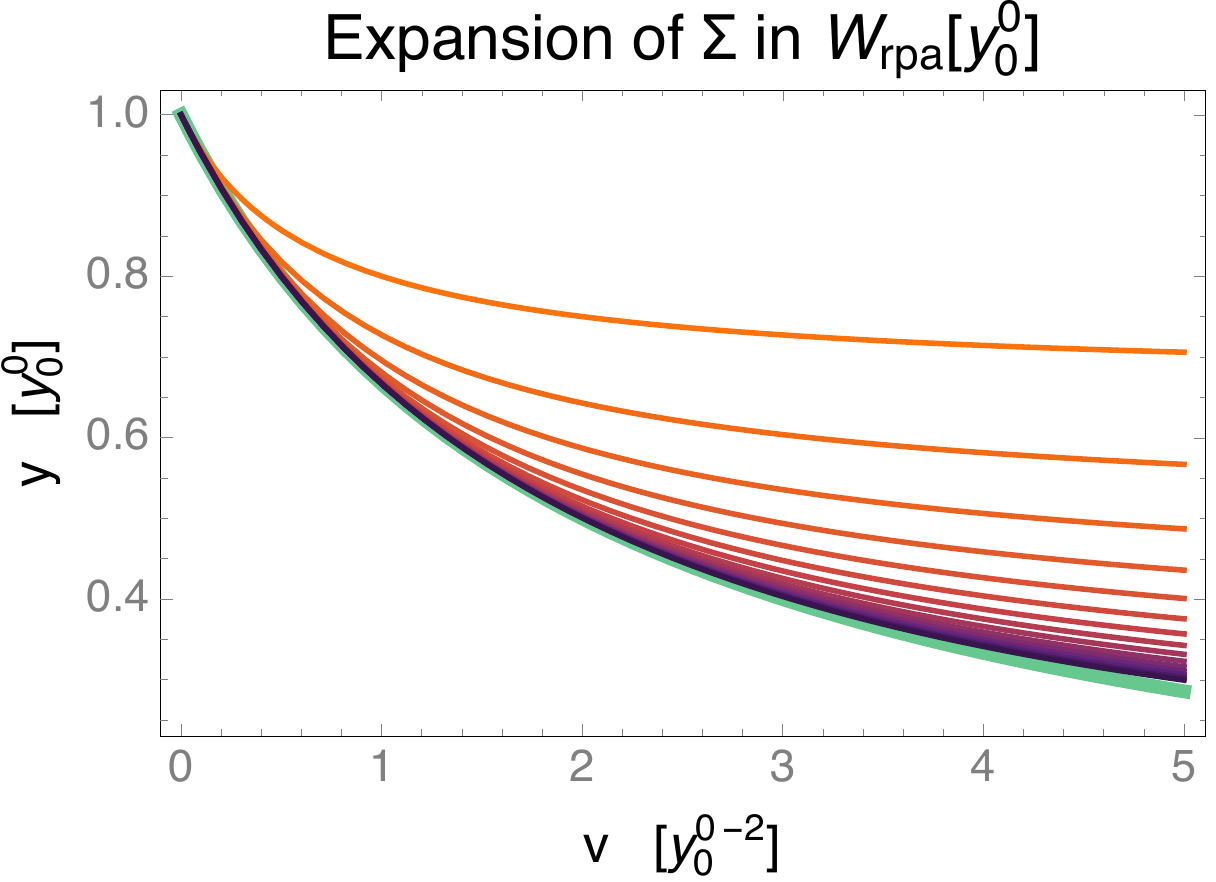}
\includegraphics[scale=0.4]{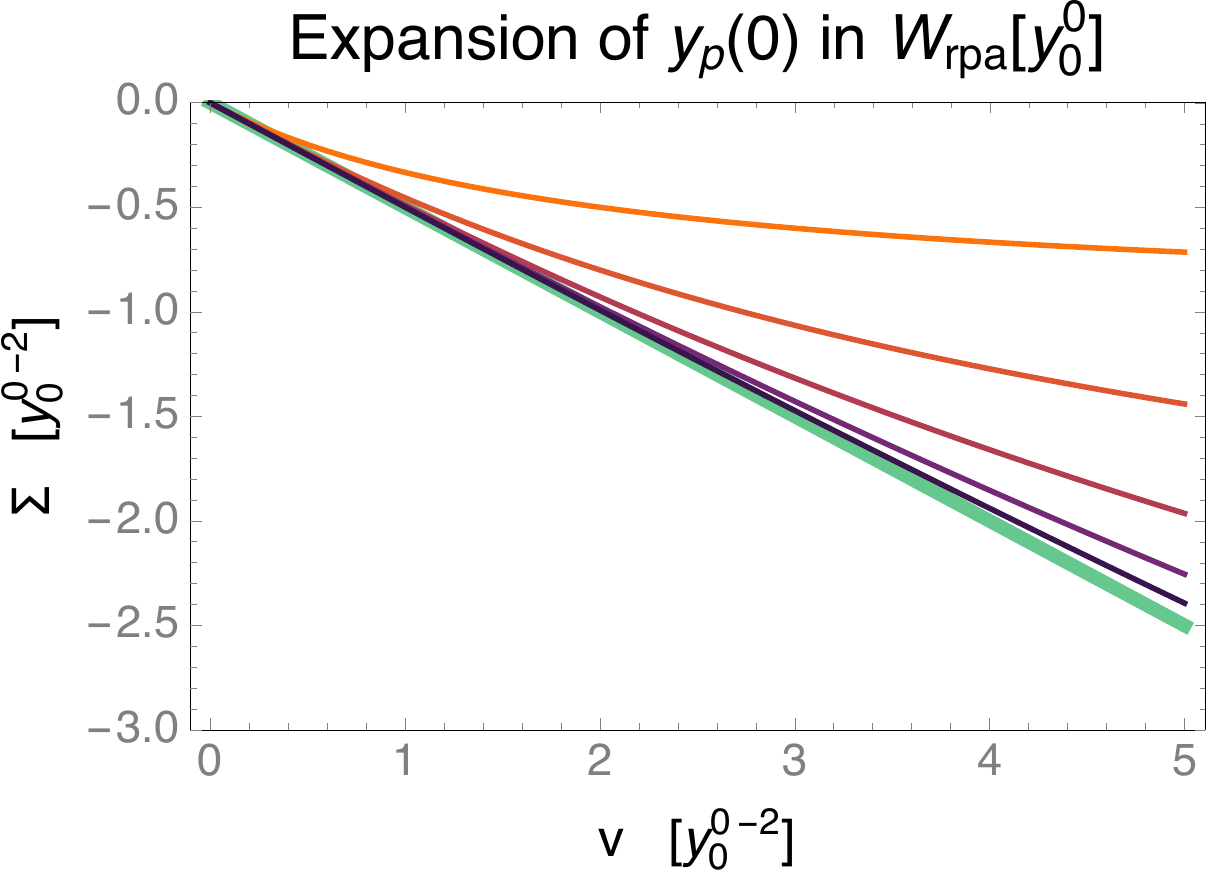}
\includegraphics[scale=0.4]{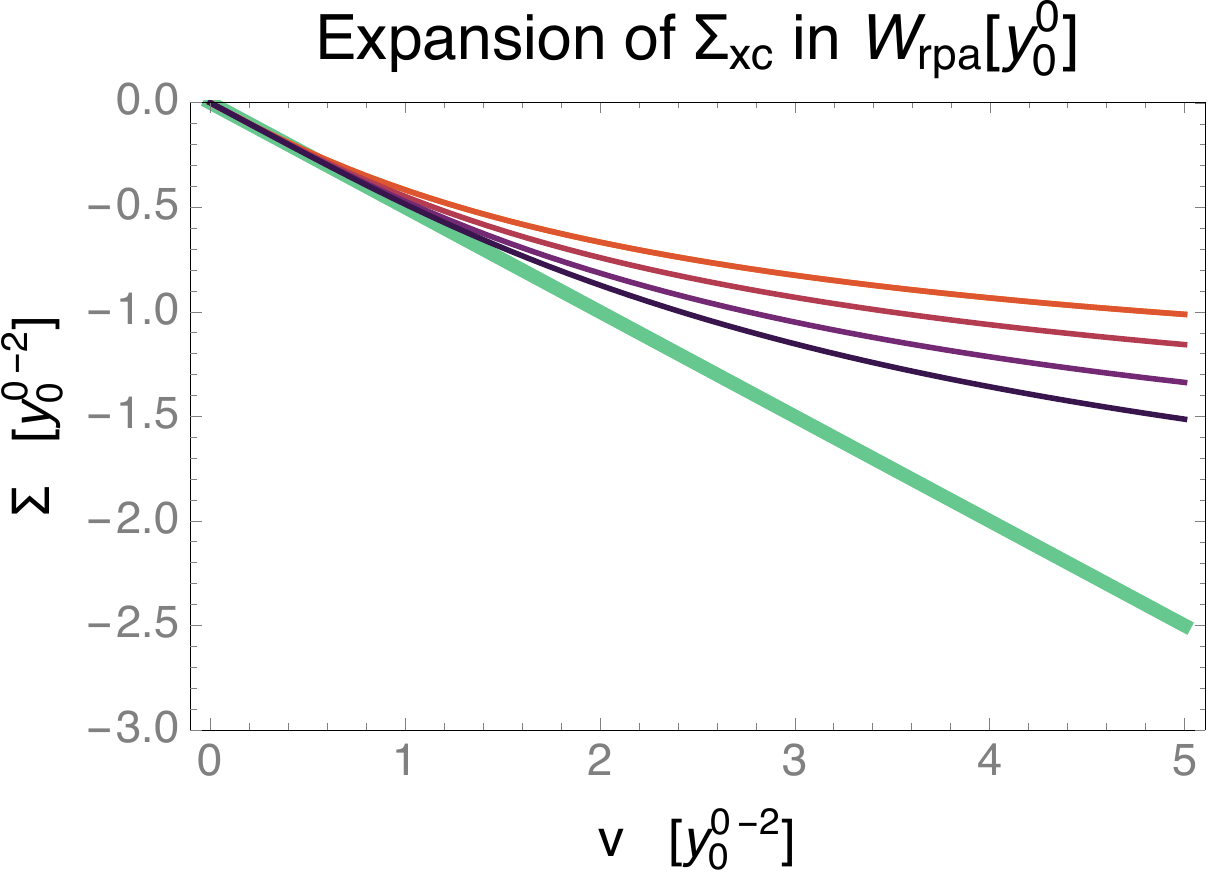}
\includegraphics[scale=0.4]{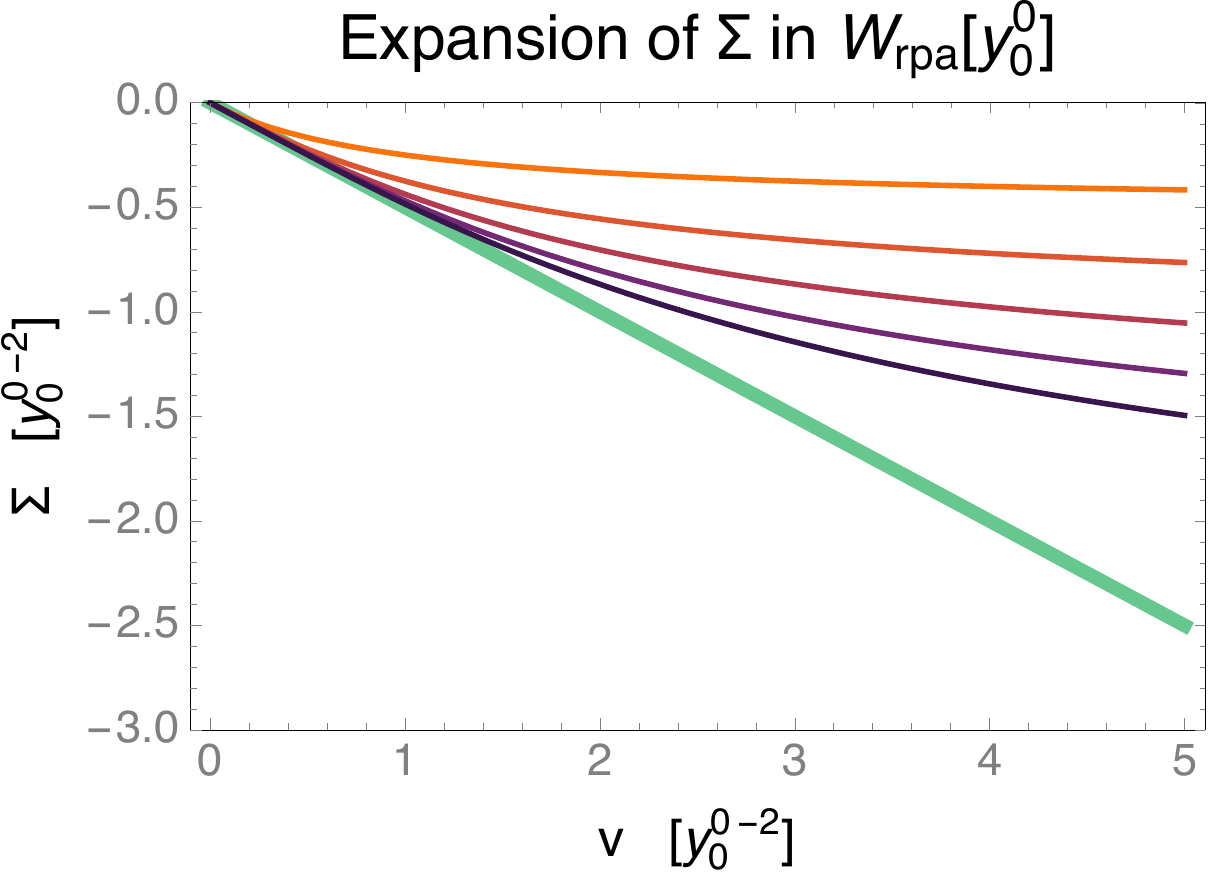}
\caption{Top panel:
The exact $y_p(0)$, in green, and the hierarchy of approximations obtained by using truncations of the series in $w_{rpa}$, given by equations (45), (46) and (47), respectively, in shades of orange (the darker the shade the higher the order of the truncation). Bottom panel:
The exact self-energy $\Sigma$, in green, and the self-energies corresponding to the above approximations, in the same shades of orange.}
\label{fig:pertrpa}
\end{figure}

The function $y_p(0)$, which represent the equilibrium interacting Green's function,
is written in terms of two parameters: $\y00$, which represents the equilibrium non-interacting 
Green's function, and $v$, which parametrizes the interaction.
Following Section \ref{sec:pertW0}, we can rewrite it in terms of $\y00$ and $w_{rpa}\equiv W_{rpa}(\y00)$
by simply using the definition of $w_{rpa}$ to express $v$, which leads to
\be\label{ywrpa}
y_p(0)=\frac{y_0^0}{1-\frac{y_0^0{}^2 w_{\text{rpa}}}{2 \left(y_0^0{}^2 w_{\text{rpa}}-1\right)}}=y_0^0-\frac{y_0^0{}^3}{2}w_{\text{rpa}}-\frac{1}{4} y_0^0{}^5 w_{\text{rpa}}^2-\frac{1}{8} y_0^0{}^7 w_{\text{rpa}}^3+...
\ee
Similarly, we can express $\Sigma$ and $\sxc$ as
\bea\label{sigmawrpa}
\Sigma&=&\frac{y_0^0 w_{\text{rpa}}}{2 y_0^0{}^2 w_{\text{rpa}}-2}=-\frac{y_0^0 }{2}w_{\text{rpa}}-\frac{1}{2} y_0^0{}^3 w_{\text{rpa}}^2-\frac{1}{2} y_0^0{}^5 w_{\text{rpa}}^3-\frac{1}{2} y_0^0{}^7 w_{\text{rpa}}^4+	...\\
\sxc&=&\frac{y_0^0 w_{\text{rpa}} \left(2-3 y_0^0{}^2 w_{\text{rpa}}\right)}{2 y_0^0{}^2 w_{\text{rpa}} \left(y_0^0{}^2 w_{\text{rpa}}-3\right)+4}=\frac{y_0^0 }{2}w_{\text{rpa}}-\frac{1}{4} y_0^0{}^5 w_{\text{rpa}}^3-\frac{3}{8} y_0^0{}^7 w_{\text{rpa}}^4+...\label{sxcg0w0}
\eea
In figure \ref{fig:pertrpa}, we compare approximations to $y_p(0)$
resulting from the straight perturbative expansion \eqref{ywrpa} and from the 
solution to the Dyson equation with self-energy approximated by the expansion in \eqref{sigmawrpa}.
All hierarchies of approximations smoothly converge to the correct result.
It is worth noticing that, even though the leading order expansion of $\sxc$ gives a better
approximation for large values of $v$ compared with the other leading orders,
the straight expansion of $y_p(0)$ converges much faster than the others.

\begin{figure}[h]
\includegraphics[scale=0.7]{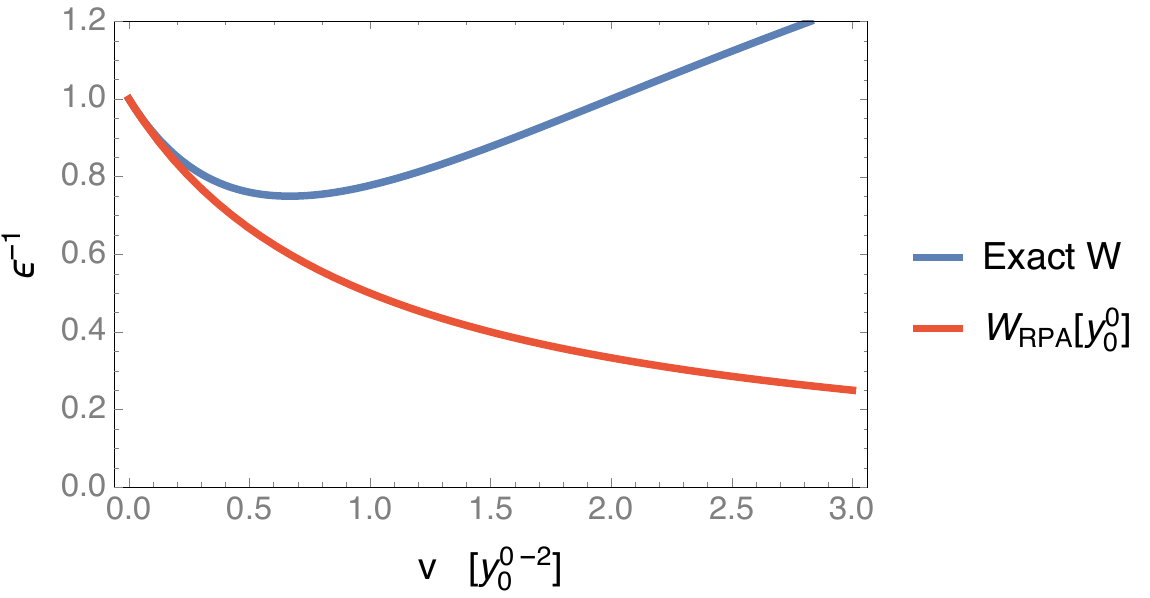}
\caption{The exact inverse of the dielectric function, defined as $\epsilon^{-1}\equiv W/v$, compared to its RPA equivalent $\epsilon^{-1}_{RPA}\equiv W_{rpa}(y^0_0)/v$.}
\label{fig:epsm1}
\end{figure}

\begin{figure}[h]
\includegraphics[scale=0.4]{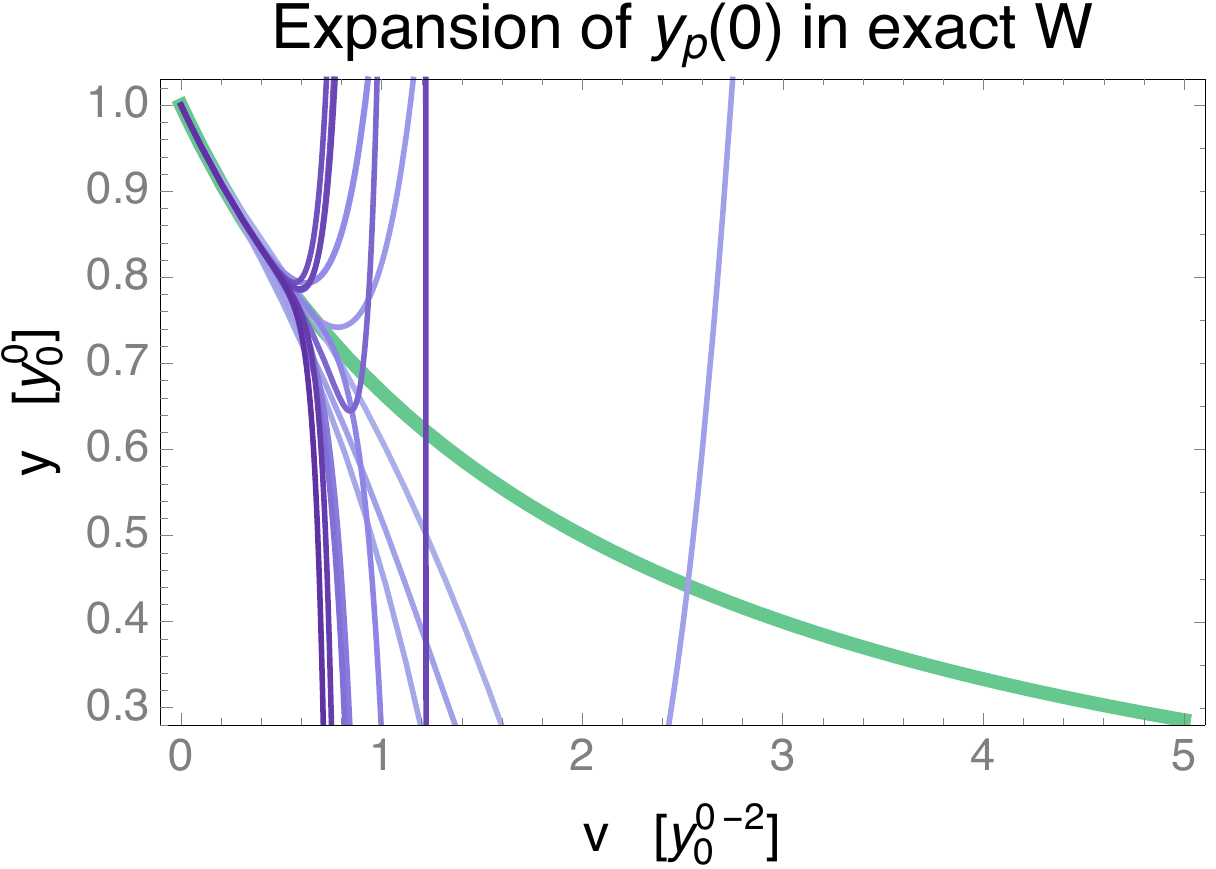}
\includegraphics[scale=0.4]{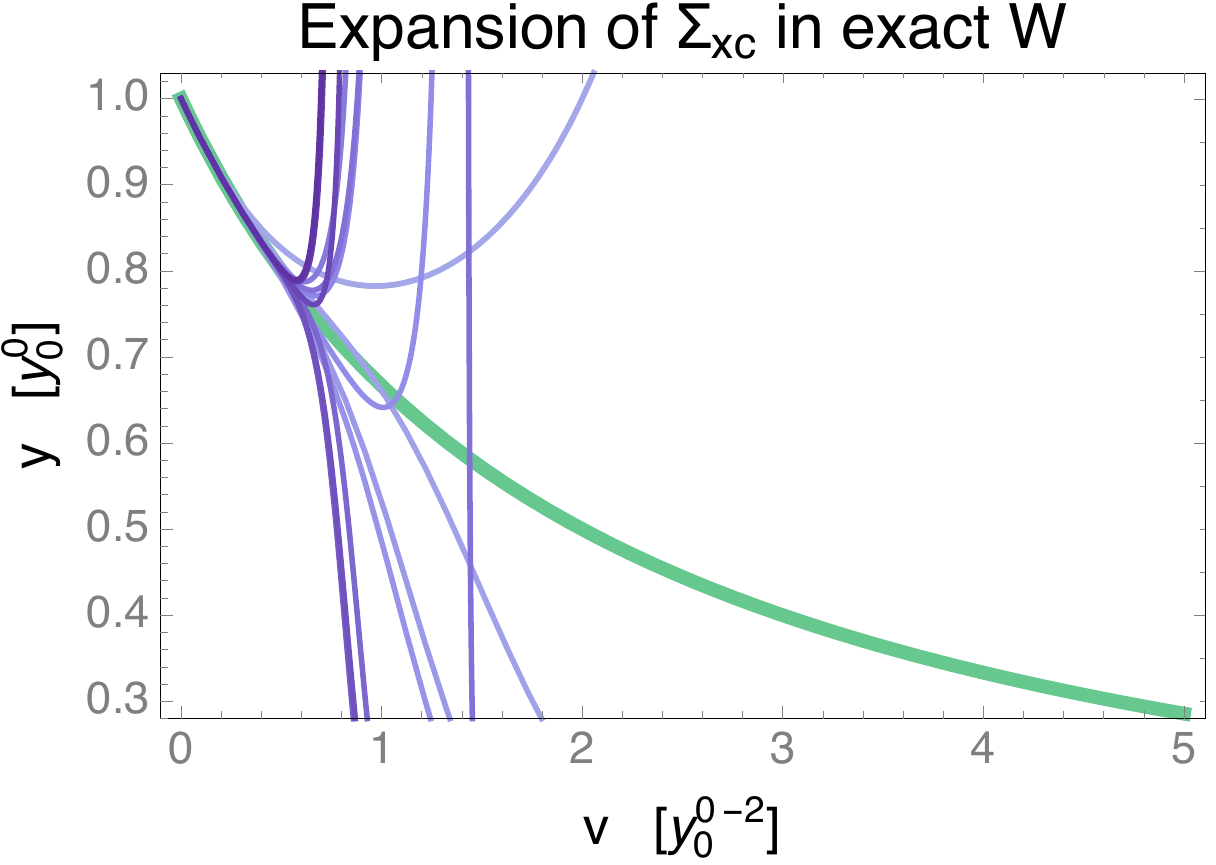}
\includegraphics[scale=0.4]{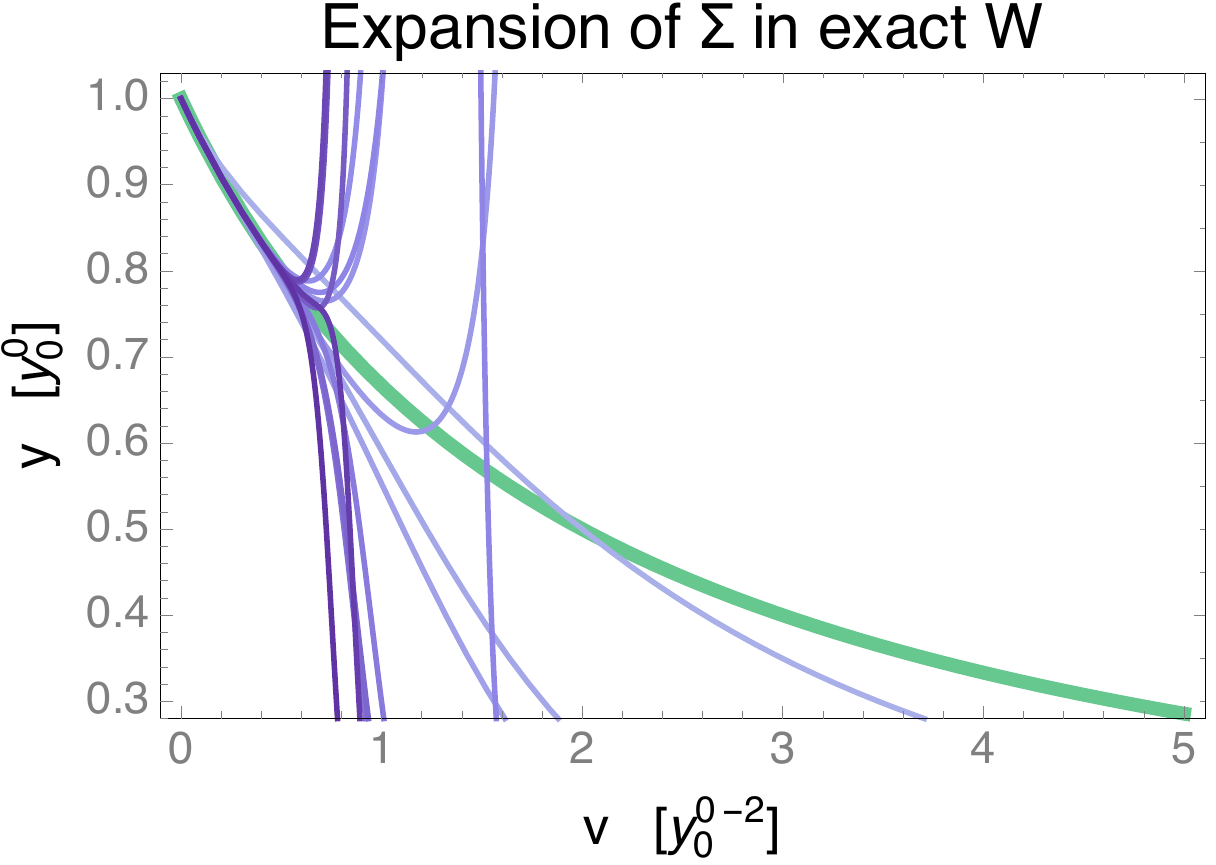}
\caption{The exact $y_p(0)$, in green, and the hierarchy of approximations obtained by using trunctions of the series in the exact screening $W$, given by eqns (49), (50) and (51), respectively, in shades of blue (the darker the shade the higher the order of the truncation).}
\label{fig:pertw}
\end{figure}

Finally, we would like to reply to the question: is $w_{rpa}$ a better perturbative
parameter than the exact $W$? To do that, one has to invert the expression \eqref{wrpadef}
to get the function corresponding to the map $W \to v$. There are three solutions to the equation for $v$, of which only one goes to 0 when $W$
goes to 0, namely
\be
v=\frac{y_0^0{}^2 w \left(\sqrt[3]{36 \sqrt{3} A+B}+y_0^0{}^2 w+36\right)+\left(36 \sqrt{3} A+B\right)^{2/3}-36}{9 y_0^0{}^2 \sqrt[3]{36 \sqrt{3} A+B}}
\ee
with $A\equiv \sqrt{y_0^0{}^2 w \left(y_0^0{}^2 w \left(2 y_0^0{}^2 w+83\right)-36\right)+12}$
and $B\equiv w y_0^0{}^2 (432 + w y_0^0{}^2 (54 + w y_0^0{}^2))$.
Provided with such a map, we can calculate again the expansion of $y_p(0)$, $\sxc$ and $\Sigma$
to be
\bea\label{sigmawexact}
y_p(0)&=& y_0^0-\frac{y_0^0{}^3 w}{2}-\frac{y_0^0{}^5 w^2}{4}+\frac{y_0^0{}^7 w^3}{8}+\frac{13 y_0^0{}^9 w^4}{16}+\frac{43 y_0^0{}^{11} w^5}{32}+...\\
\Sigma&=&-\frac{y_0^0 w}{2}-\frac{y_0^0{}^3 w^2}{2}-\frac{y_0^0{}^5 w^3}{4}+\frac{5 y_0^0{}^7 w^4}{8}+\frac{33 y_0^0{}^9 w^5}{16}+...\\
\sxc&=&\frac{y_0^0 w}{2}-\frac{y_0^0{}^5 w^3}{2}-y_0^0{}^7 w^4-\frac{5 y_0^0{}^9 w^5}{8}+...
\eea
Truncations of the above series are plotted in figure \ref{fig:pertw}.
Few orders of the series are representative of the overall bad behaviour of the series.
This suggests that the exact $W$ does not represent a good parameter for a perturbation expansion. Indeed, a glance at the comparison of the RPA and the exact inverse dielectric function in fig. \ref{fig:epsm1} shows 
that while the RPA value is steadily decreasing with increasing interaction, the tendency is inverted starting from a critical value of the interaction in the exact case, making the screened interaction too large to be a useful parameter for perturbation theory.

\subsection{Solutions to the Hierarchy of linearized KBE}\label{subsec:nplr}

We will now move on to the second route beyond the standard approximations, namely the expansion of the Hartree potential introduced in Section \ref{sec:nonepertlinear}. To this end, we define  
the 
the \emph{`perturbed' Hartree potential} as $V_H(x)\equiv -vy(x)$, 
which leads to rewrite \eqref{opmkbe} as
\be\label{opmkbe2}
y(x)=\y00 +\y00 \left( x+V_H(x)+\frac{1}{2}vy'(x)y(x)^{-1} \right) y(x),
\ee
and consider the hierarchy of approximations generated by truncating its
Taylor expansion in the perturbing potential
\be\label{vhseries}
V_H(x)=V_0+V_1 x+V_2 x^2+...\;.
\ee
The leading term $V_0$ simply corresponds to the $\sh$ previously defined and assumed to be known, 
while the other terms involve derivatives of $y(x)$ evaluated at zero:
$V_1=-vy'(0)$, $V_2=-vy''(0)/2$, \textit{etc.} which are not known \textit{a priori}.
Just like for $V_0$, we could consider them as given inputs, since for real systems
they can in principle be obtained within the framework of TDDFT.
In practice, however, we can only get approximations to their value, on which we have no control.
We therefore prefer to calculate them self-consistently within our scheme.
To simplify this task, we express them in terms of $\y00$, $v$ and the unknown $y(0)$, 
which is achieved by setting $x$ to 0 in the equation \eqref{opmkbe2} and its derivatives.
This leads to:
\beas\label{vhy0}
V_1&=\frac{2 y_0^0{}^3 v-6 y_0^0{}^2 v y(0)+4 y_0^0-4 y(0)}{y_0^0{}^3 v+2 y_0^0}\\ \label{vhy1}
V_2&=\frac{y_0^0 \left(v \left(3 y_0^0 y(0) \left(y_0^0{}^2 v \left(y_0^0{}^2 v-2\right)-4\right)-2 y(0)^2 \left(y_0^0{}^2 v+2\right) \left(3 y_0^0{}^2
   v+2\right)+2 y_0^0{}^2 \left(3 y_0^0{}^2 v+8\right)\right)+8\right)-8 y(0)}{y_0^0{}^2 v \left(y_0^0{}^2 v+2\right){}^2}\\
&...
\end{align}\end{subequations}

In previous works \cite{lani2012,berger2014}, the zeroth and first order approximations to $V_H$,
namely $V_H\approx V_0$ and $V_H(x)\approx V_0+xV_1$, were considered,
although, as pointed out earlier, in slightly different versions from what we are now studying.\footnote{%
Lani \textit{et al.} considered $V_H(x)\approx V_0$ but in the OPM with $\lambda=1$,
while Berger \textit{et al.} considered $V_H(x)\approx V_0+x V_1$ but with $V_0$
self-consistently determined.}

We will now establish a general procedure that solves the equation for an arbitrarily high order 
truncation of \eqref{vhseries}. We start by noticing that, since $V_H(x)$ is a given function,
the equation \eqref{opmkbe2} is in the linear form
\be\label{lindiffeqn}
y'(x)=\alpha(x)+\beta(x) y(x).
\ee
For $\alpha(x)=0$, the family of solutions of the above equation is
\be
y(x)=Ce^{\int \beta(x) dx}.
\ee
This suggests to use the ansatz $y(x)=\bar{y}(x)e^{\int \beta(x) dx}$ when $\alpha(x)\neq 0$.
When inserted in \eqref{lindiffeqn}, we obtain the following equation for $\bar{y}(x)$:
\be
\bar{y}'(x)=\alpha(x)e^{-\int \beta(x) dx},
\ee
which is solved by the family of functions
\be
\bar{y}_C(x)=C+\int_a^x \alpha(x')e^{-\int \beta(x') dx'}dx',
\ee
where the lower bound is a fixed constant, while $C$ parametrizes the family, leading to
\be\label{sollineqdiff}
y_C(x)=e^{\int \beta(x) dx}\left(C+\int_a^x \alpha(x')e^{-\int \beta(x') dx'}dx'\right).
\ee
The value of $a$ must be chosen carefully once the integrand $\beta(x)$ is specified, in order to avoid possible divergences of the integral.
When applied to \eqref{opmkbe2}, we have the exact result
\be
\beta(x)=-\frac{2 (y_0^0 {V_H}(x)+y_0^0 x-1)}{y_0^0 v}\;\;\;\;\mbox{and}\;\;\;\;\alpha(x)=-\frac{2}{v}.
\ee
Since $a$ has the dimension of a potential, we chose  $a=\y00{}^{-1}$,
which will turn out to create no problems of convergence of the integral.
We finally arrive at
\be\label{opmansatz}
y_C(x)=e^{-f(x)}\left(C-\frac{2}{v}\int_{\y00{}^{-1}}^{x}dx'e^{f(x')}\right)
\ee
with	
\be
f(x)=\int dx\left(\frac{2 \left(y_0^0 \left(V_0+V_1 x+V_2 x^2+...\right)+y_0^0 x-1\right)}{y_0^0 v}\right)
=\frac{2 \left(y_0^0 V_0 x+\frac{1}{2} y_0^0 V_1 x^2+\frac{1}{3} y_0^0 V_2 x^3+...+\frac{y_0^0 x^2}{2}-x\right)}{y_0^0 v}.
\ee
which is the family of solutions to \eqref{opmkbe2}, parametrized by the constant $C$.

Now that the family of solutions is established, the next task is to identify the physical one.
It is possible to prove that the choice $C=0$ is sufficient to pick a solution that reproduces the correct noninteracting
function $y_0(x)$ in the limit $v\to 0$, therefore we shall refer to that as the 
\emph{approximate physical solution}.
It should be noticed that $C=0$, however, is not the only possible choice.
This is because $C$ is a constant with respect to $x$, the variable in the differential problem \eqref{opmkbe2},
but it can still depend on $v$ and $y_0^0$ via the dimensionless quantity $v\y00{}^2$;
it follows that any choice of such a dependence such that $e^{-f(x)}C\to 0$ for $v\to 0$ would be allowed.
This is a fundamental problem of the framework set by the Kadanoff-Baym equation,
which will not be addressed here, but is left to future work.
For the moment it should suffice to say that an arbitrary choice of such a dependency
identifies a hierarchy of approximate physical solutions that do not, in general, 
converge to the exact one; our choice $C=0$, on the other hand, seems to
create a hierarchy with the correct convergence. Although we do not have an analytic proof for this fact,
we do have strong numerical evidence, as we shall soon see.

Since each truncation of \eqref{vhseries}
corresponds to a different approximation of the original equation \eqref{opmkbe2},
we label the corresponding physical solution with an index $(n)$, where $n$ is the order of the truncation:
\begin{multline}\label{opmypnx}
y^{(n)}_p(x)=-\frac{2}{v}
	e^{-\frac{2 \left(y_0^0 V_0 x+\frac{1}{2} y_0^0 V_1 x^2+\frac{1}{3} y_0^0 V_2 x^3+...+\frac{1}{n+1} y_0^0 V_n x^{n+1}+\frac{y_0^0 x^2}{2}-x\right)}{y_0^0 v}}\times\\
\times\int_{\y00{}^{-1}}^{x}dx'
	e^{\frac{2 \left(y_0^0 V_0 x'+\frac{1}{2} y_0^0 V_1 x'^2+\frac{1}{3} y_0^0 V_2 x'^3+...+\frac{1}{n+1} y_0^0 V_n x'^{n+1}+\frac{y_0^0 x'^2}{2}-x'\right)}{y_0^0 v}},
\end{multline}
which at equilibrium simplifies to
\be\label{opmypn}
y^{(n)}_p(0)=\y00\int_{0}^{1}\!d\xi\; \frac{2}{v\y00{}^2}\;
	e^{\frac{2 \left(y_0^0 V_0 \xi+\frac{1}{2} V_1 \xi^2+\frac{1}{3} y_0^0{}^{-1} V_2 \xi^3+...+\frac{1}{n+1} y_0^0{}^{1-n} V_n \xi^{n+1}+\frac{ \xi^2}{2}-\xi\right)}{v\y00{}^2}}.
\ee
Exact expressions for the integral in \eqref{opmypn} can be found for $n=0$ and $n=1$,
while for higher orders the integral can be evaluated numerically.

\begin{figure}[h]
\includegraphics[scale=0.4]{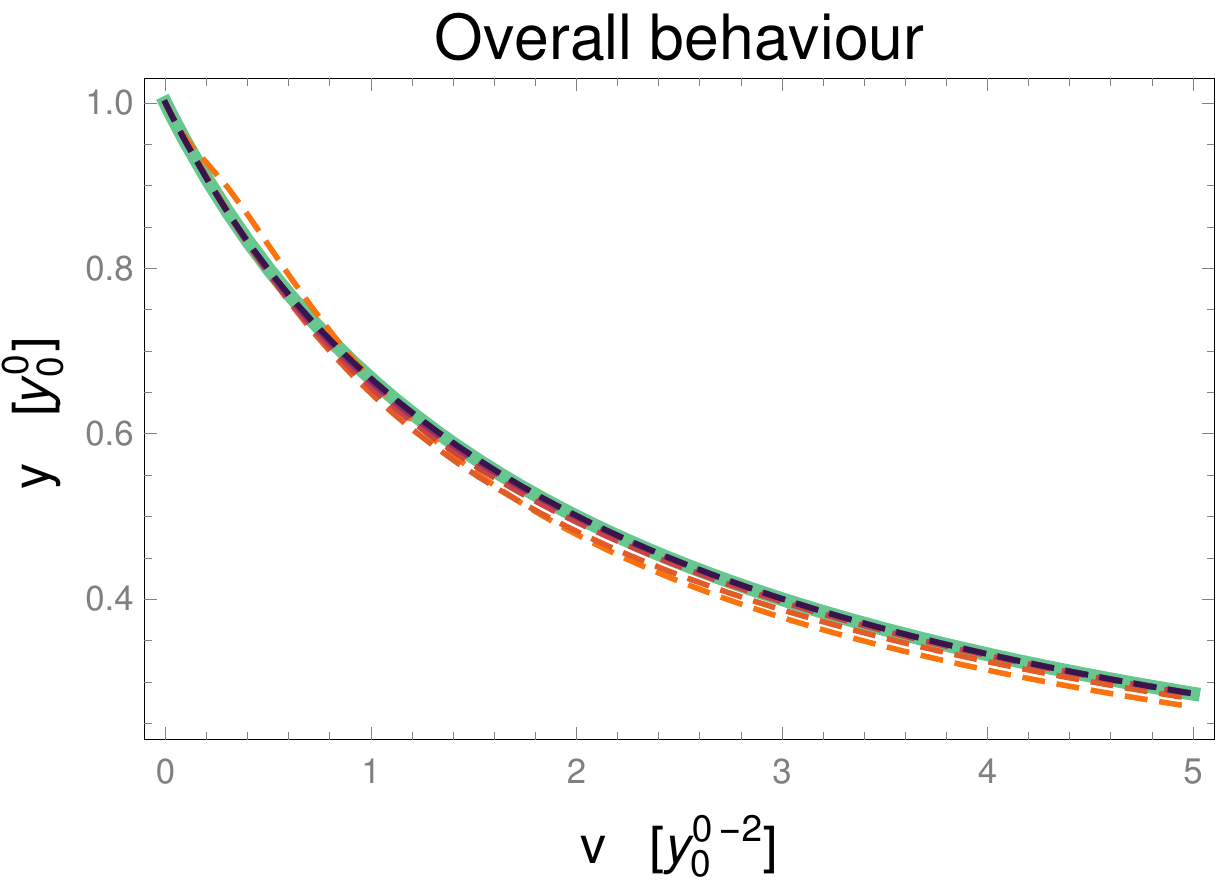}
\includegraphics[scale=0.4]{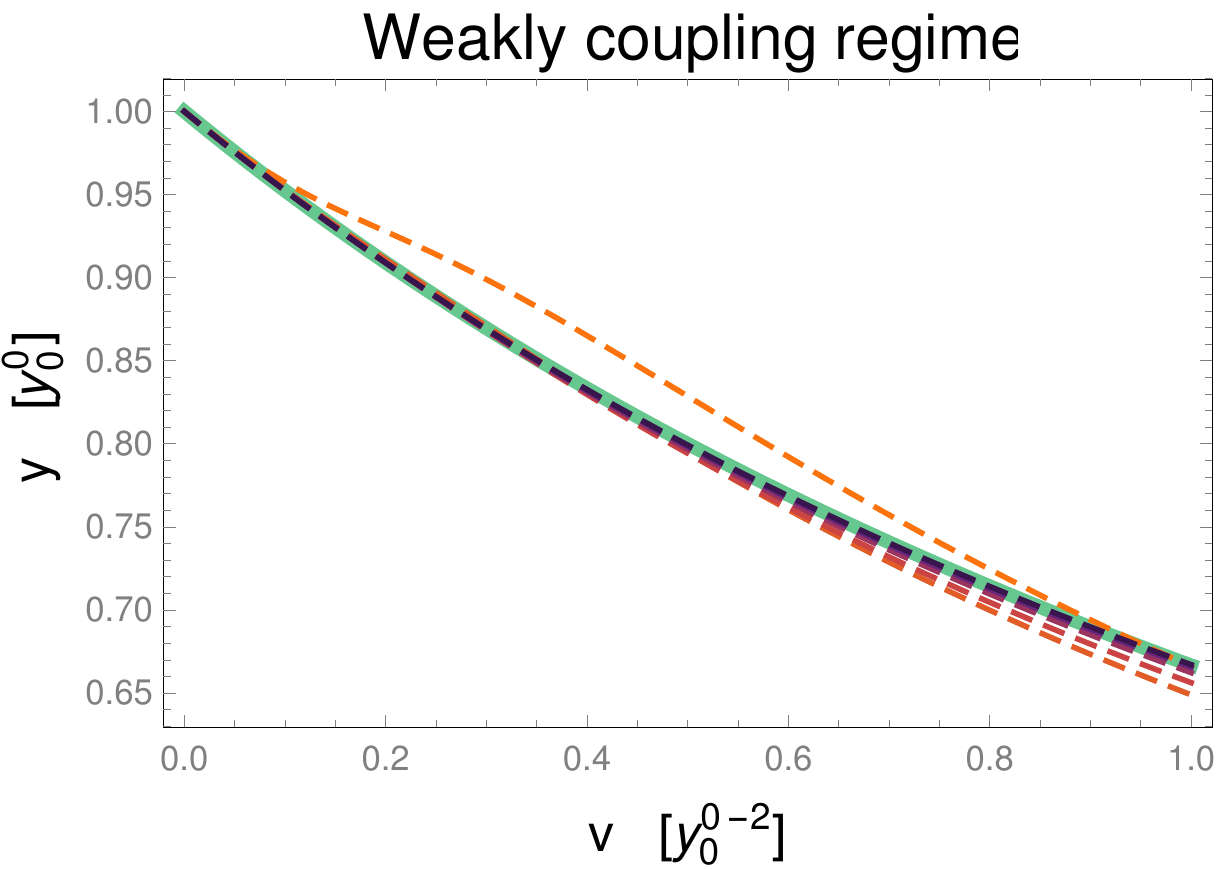}
\includegraphics[scale=0.4]{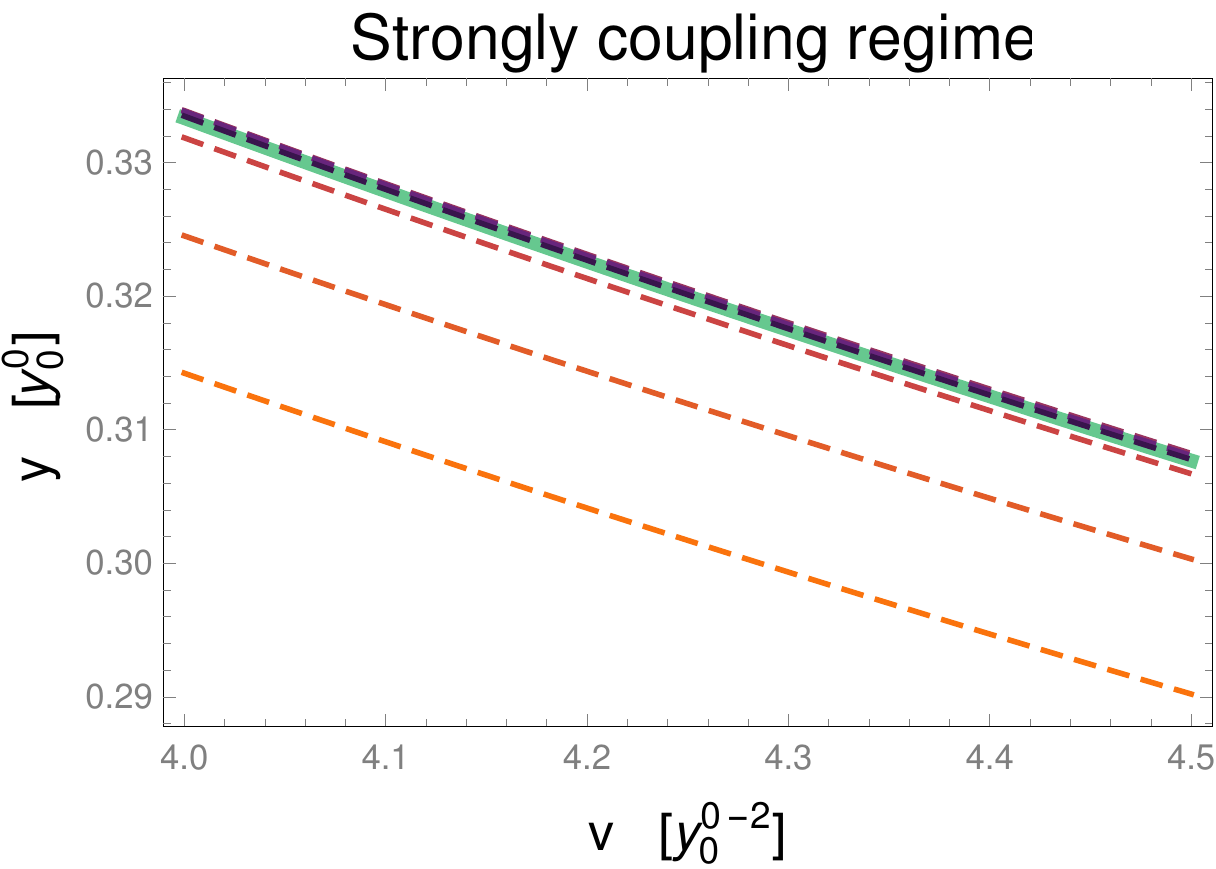}
\includegraphics[scale=0.5]{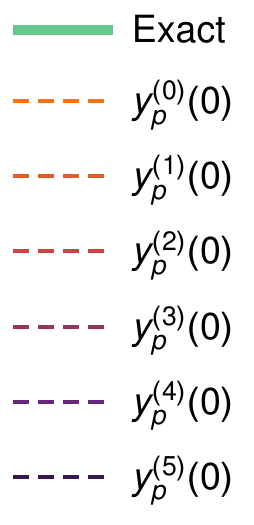}
\caption{The physical solutions to \eqref{opmkbe2} with approximate $V_H(x)$
are compared with the exact physical solution $y_p(0)$. The function $y^{(0)}_p(0)$
corresponds to $V_H\approx V_0$, $y^{(1)}_p(0)$ to $V_H\approx V_0+x V_1$ and so on.}
\label{fig:y0123}
\end{figure}

As a matter of fact, equation \eqref{opmypn} only gives an implicit definition of $y^{(n)}_p(0)$,
since $V_1$, $V_2$, ... also depend on $y^{(n)}_p(0)$ via (\ref{vhy0},\ref{vhy1},...) where $y(0)$
has to be replaced by $y^{(n)}_p(0)$. Although already for $n=1$ we cannot 
find a closed, explicit form for $y^{(n)}_p(0)$, 
these functions can be evaluated numerically.
\footnote{%
We would like to point out that there is strong numerical evidence that these equations, 
although highly nonlinear, seem to have only one solution,
so there is no problem of multiple, spurious solutions at this level.}
In figure \ref{fig:y0123} the first six functions $y^{(n)}_p(0)$, namely $n=0,1,2,3,4,5$,
are shown as function of $v$ in comparison with the exact physical solution $y_p(0)$ (eq. \ref{opmypn}).

\subsection{Comparing all approximations}\label{subsec:comp}
\begin{figure}[h]
\includegraphics[scale=0.8]{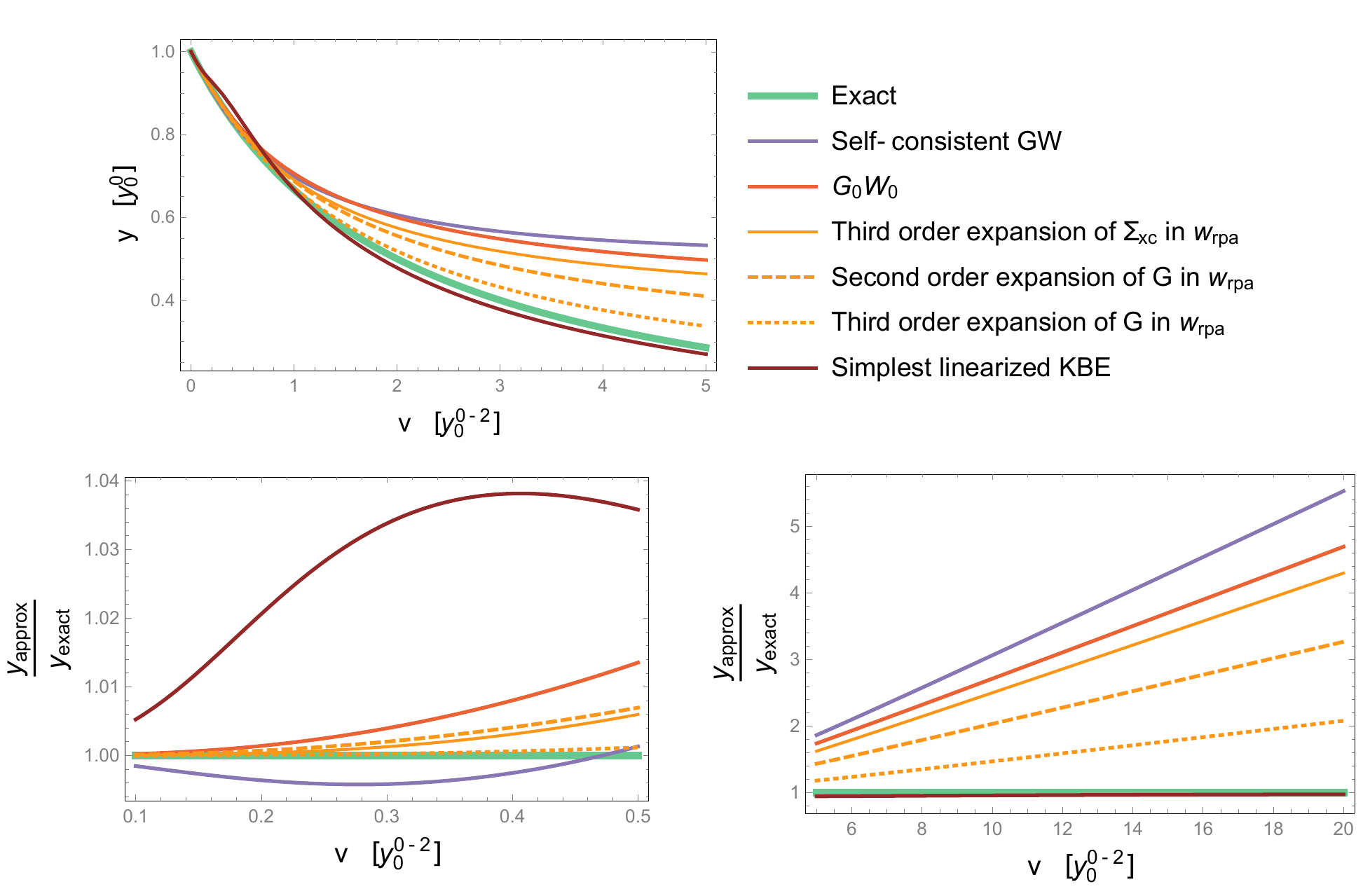}
\caption{We here compare all approximations to the exact function \eqref{exacty}: the self-consistent GW \eqref{ygw}; $G_0W_0$, which is obtained by taking only the leading order in 
 \eqref{sxcg0w0}; the next order expansion of $\Sigma_{\rm xc}$ in $W_0$, obtained by taking the leading and next-to-leading order in  \eqref{sxcg0w0}; the second and third order expansion of $G$ in $W_0$ from \eqref{ywrpa}; and, finally, the physical solution to the simplest approximation to the KBE, namely $y_p^{(0)}$ from \eqref{opmypn}. Top panel: overall behavior; bottom panels: ratio between approximate and exact functions in the weakly (left) and strongly (right) coupling regimes.
}
\label{fig:compareall}
\end{figure}

A comparison between all different approximations is reported in fig. \ref{fig:compareall}.
For the expansion in $\Wrpa{G_0}$ we consider a $\sxc$ approximated with the first nontrivial 
term after the one prescribed by $G_0W_0$ (which is of third order in $\Wrpa{y_0^0}$, the second
being identically zero); we also report the second order expansion of $y(0)$ in $\Wrpa{y_0^0}$,
actually a better estimate than the expansion of $\sxc$ at lower computational cost.
For the linearized KBE, we consider the simplest approximation $V_H(x)\approx V_H(0)$.
We see that the expansion in $\Wrpa{G_0}$ leads to an improvement over $G_0W_0$, as hoped.
However, in the strongly coupled regime, 
the best approximation is given by the solution to the linearized KBE $y^{(0)}_p(0)$.

\section{Conclusions}\label{sec:conclusions}

We presented two approaches aimed at improving on the GWA outside of the framework set by 
Hedin in his original formulation. 
The first one relies on the fact that $G_0W_0$ can be regarded as
the leading order expansion of $\sxc$ in $G_0$ and $W_0\equiv \Wrpa{G_0}$.
Since there are reasons to believe that $W_0$ 
may be a better parameter for such a perturbative expansion compared to 
the bare Coulomb potential $v_c$ and even the exact screened one $W$,
we then propose to consider higher order terms in such an expansion.

The second approach relies on a Taylor expansion of the Hartree potential
in terms of the perturbing potential $V_{\rm H}[\varphi]$ in the KBE. We consider
the hierarchy of linearized KBEs arising by higher and higher order approximations
to $V_{\rm H}[\varphi]$.

Both schemes have been tested on the one-point model. This is not a `model system',
in the sense of a model for physical systems, but rather as a `model functional',
in the sense of a model for the functionals used in the Many-body Perturbation Theory.
The OPM allows to avoid certain theoretical limitations of our current perturbative tools
and explore aspects of the theory that would be computationally too demanding to study in real systems.
This is a particularly suitable framework for our study, since,
as mentioned, there is not a known systematic way to deal with 
equations such as the KBE.
Nonetheless, the procedure we used for the OPM
is a necessary proof of principle of the feasibility of the approach
and gives leads on how to tackle the full many-body problem.

Results concerning the quality of the approximations in comparison with the OPM equivalent of GWA are encouraging.
The series in $\Wrpa{G_0}$ seems to be very well behaved
and provide good estimates of Green's functions in the strong coupling regimes
already with a few orders. This contrasts with the expansion in $W$,
which presents quite an erratic behaviour.
The second approach works better, in fact beyond expectations.
The solution to even the simplest linearized KBE leads to an approximate 
Green's function that outperform any other perturbative approach in the strong coupling regime.
This suggests that the nonperturbative character of the KBE
makes the equation a particularly promising starting point for 
methods in the strong coupling regime.

For a real system, the first approach presents a theoretically clear path to follow.
The way to build the hierarchy of approximations is well defined and,
being an expansion in the bare Green's function, there are no problems of spurious solutions.
The only limitation is the computational cost of the approach, which is the same as that of 
an expansion in $G_0$ and $v_c$.
The second approach is more sophisticated from a mathematical perspective, 
but at the same time more clear from a physics point of view: 
the only approximation is made in the response of the Hartree
potential to a perturbing potential. However, the approach is theoretically incomplete,
for a lack of mathematical tools for a systematic treatment of the 
pertinent equations. Our findings in the OPM may give us some guidance,
but a complete picture of the procedure for a real system is still to be achieved.
Nonetheless, our preliminary tests show such a striking improvement compared 
with competing perturbative methods that  more work in this direction is certainly motivated.

\section{Acknowledgements}
The research leading to these results has received
funding from the European Research Council under the European Union’s Seventh Framework Programme
(FP/2007–2013) / ERC Grant Agreement n 320971.
BSM acknowledges the Laboratoire des Solides Irradi\'es (Ecole
Polytechnique, Palaiseau, France) for the support and hospitality
during a sabbatical year. BSM acknowledges partial support from
CONACYT-M\'exico Grant 153930.


\bibliography{ref}

\begin{thebibliography}{18}
\expandafter\ifx\csname natexlab\endcsname\relax\def\natexlab#1{#1}\fi
\expandafter\ifx\csname bibnamefont\endcsname\relax
  \def\bibnamefont#1{#1}\fi
\expandafter\ifx\csname bibfnamefont\endcsname\relax
  \def\bibfnamefont#1{#1}\fi
\expandafter\ifx\csname citenamefont\endcsname\relax
  \def\citenamefont#1{#1}\fi
\expandafter\ifx\csname url\endcsname\relax
  \def\url#1{\texttt{#1}}\fi
\expandafter\ifx\csname urlprefix\endcsname\relax\def\urlprefix{URL }\fi
\providecommand{\bibinfo}[2]{#2}
\providecommand{\eprint}[2][]{\url{#2}}

\bibitem[{\citenamefont{Schwinger}(1951{\natexlab{a}})}]{schwinger1951a}
\bibinfo{author}{\bibfnamefont{J.}~\bibnamefont{Schwinger}},
  \bibinfo{journal}{Proceedings of the National Academy of Sciences}
  \textbf{\bibinfo{volume}{37}}, \bibinfo{pages}{452}
  (\bibinfo{year}{1951}{\natexlab{a}}),
  \eprint{http://www.pnas.org/content/37/7/452.full.pdf},
  \urlprefix\url{http://www.pnas.org/content/37/7/452.short}.

\bibitem[{\citenamefont{Schwinger}(1951{\natexlab{b}})}]{schwinger1951b}
\bibinfo{author}{\bibfnamefont{J.}~\bibnamefont{Schwinger}},
  \bibinfo{journal}{Proceedings of the National Academy of Sciences}
  \textbf{\bibinfo{volume}{37}}, \bibinfo{pages}{455}
  (\bibinfo{year}{1951}{\natexlab{b}}),
  \eprint{http://www.pnas.org/content/37/7/455.full.pdf},
  \urlprefix\url{http://www.pnas.org/content/37/7/455.short}.

\bibitem[{\citenamefont{Hedin}(1965{\natexlab{a}})}]{hedin1965}
\bibinfo{author}{\bibfnamefont{L.}~\bibnamefont{Hedin}},
  \bibinfo{journal}{Phys. Rev.} \textbf{\bibinfo{volume}{139}},
  \bibinfo{pages}{A796} (\bibinfo{year}{1965}{\natexlab{a}}),
  \urlprefix\url{https://link.aps.org/doi/10.1103/PhysRev.139.A796}.

\bibitem[{\citenamefont{Martin et~al.}(2016)\citenamefont{Martin, Reining, and
  Ceperley}}]{the_book}
\bibinfo{author}{\bibfnamefont{R.}~\bibnamefont{Martin}},
  \bibinfo{author}{\bibfnamefont{L.}~\bibnamefont{Reining}}, \bibnamefont{and}
  \bibinfo{author}{\bibfnamefont{D.}~\bibnamefont{Ceperley}},
  \emph{\bibinfo{title}{Interacting Electrons: Theory and Computational
  Approaches}} (\bibinfo{publisher}{Cambridge University Press},
  \bibinfo{year}{2016}), ISBN \bibinfo{isbn}{9781316558560}.

\bibitem[{\citenamefont{Springer et~al.}(1998)\citenamefont{Springer,
  Aryasetiawan, and Karlsson}}]{springer1998}
\bibinfo{author}{\bibfnamefont{M.}~\bibnamefont{Springer}},
  \bibinfo{author}{\bibfnamefont{F.}~\bibnamefont{Aryasetiawan}},
  \bibnamefont{and} \bibinfo{author}{\bibfnamefont{K.}~\bibnamefont{Karlsson}},
  \bibinfo{journal}{Phys. Rev. Lett.} \textbf{\bibinfo{volume}{80}},
  \bibinfo{pages}{2389} (\bibinfo{year}{1998}),
  \urlprefix\url{https://link.aps.org/doi/10.1103/PhysRevLett.80.2389}.

\bibitem[{\citenamefont{Kadanoff et~al.}(1994)\citenamefont{Kadanoff, Baym, and
  Pines}}]{kadanoff}
\bibinfo{author}{\bibfnamefont{L.}~\bibnamefont{Kadanoff}},
  \bibinfo{author}{\bibfnamefont{G.}~\bibnamefont{Baym}}, \bibnamefont{and}
  \bibinfo{author}{\bibfnamefont{D.}~\bibnamefont{Pines}},
  \emph{\bibinfo{title}{Quantum Statistical Mechanics}}, Advanced Books
  Classics Series (\bibinfo{publisher}{Avalon Publishing},
  \bibinfo{year}{1994}), ISBN \bibinfo{isbn}{9780201410464},
  \urlprefix\url{https://books.google.co.uk/books?id=hIzbYznK1IYC}.

\bibitem[{\citenamefont{Pavlyukh and H\"ubner}(2007)}]{pavlyukh2007}
\bibinfo{author}{\bibfnamefont{Y.}~\bibnamefont{Pavlyukh}} \bibnamefont{and}
  \bibinfo{author}{\bibfnamefont{W.}~\bibnamefont{H\"ubner}},
  \bibinfo{journal}{Journal of Mathematical Physics}
  \textbf{\bibinfo{volume}{48}}, \bibinfo{pages}{052109}
  (\bibinfo{year}{2007}), \eprint{http://dx.doi.org/10.1063/1.2728512},
  \urlprefix\url{http://dx.doi.org/10.1063/1.2728512}.

\bibitem[{\citenamefont{Molinari}(2005)}]{molinari2005}
\bibinfo{author}{\bibfnamefont{L.~G.} \bibnamefont{Molinari}},
  \bibinfo{journal}{Phys. Rev. B} \textbf{\bibinfo{volume}{71}},
  \bibinfo{pages}{113102} (\bibinfo{year}{2005}),
  \urlprefix\url{http://link.aps.org/doi/10.1103/PhysRevB.71.113102}.

\bibitem[{\citenamefont{Lani et~al.}(2012)\citenamefont{Lani, Romaniello, and
  Reining}}]{lani2012}
\bibinfo{author}{\bibfnamefont{G.}~\bibnamefont{Lani}},
  \bibinfo{author}{\bibfnamefont{P.}~\bibnamefont{Romaniello}},
  \bibnamefont{and} \bibinfo{author}{\bibfnamefont{L.}~\bibnamefont{Reining}},
  \bibinfo{journal}{New Journal of Physics} \textbf{\bibinfo{volume}{14}},
  \bibinfo{pages}{013056} (\bibinfo{year}{2012}),
  \urlprefix\url{http://stacks.iop.org/1367-2630/14/i=1/a=013056}.

\bibitem[{\citenamefont{Kozik et~al.}(2015)\citenamefont{Kozik, Ferrero, and
  Georges}}]{kozik}
\bibinfo{author}{\bibfnamefont{E.}~\bibnamefont{Kozik}},
  \bibinfo{author}{\bibfnamefont{M.}~\bibnamefont{Ferrero}}, \bibnamefont{and}
  \bibinfo{author}{\bibfnamefont{A.}~\bibnamefont{Georges}},
  \bibinfo{journal}{Phys. Rev. Lett.} \textbf{\bibinfo{volume}{114}},
  \bibinfo{pages}{156402} (\bibinfo{year}{2015}),
  \urlprefix\url{http://link.aps.org/doi/10.1103/PhysRevLett.114.156402}.

\bibitem[{\citenamefont{Tarantino et~al.}(2017)\citenamefont{Tarantino,
  Romaniello, Berger, and Reining}}]{tarantino}
\bibinfo{author}{\bibfnamefont{W.}~\bibnamefont{Tarantino}},
  \bibinfo{author}{\bibfnamefont{P.}~\bibnamefont{Romaniello}},
  \bibinfo{author}{\bibfnamefont{J.~A.} \bibnamefont{Berger}},
  \bibnamefont{and} \bibinfo{author}{\bibfnamefont{L.}~\bibnamefont{Reining}},
  \bibinfo{journal}{Phys. Rev. B} \textbf{\bibinfo{volume}{96}},
  \bibinfo{pages}{045124} (\bibinfo{year}{2017}),
  \urlprefix\url{https://link.aps.org/doi/10.1103/PhysRevB.96.045124}.

\bibitem[{\citenamefont{Stan et~al.}(2015)\citenamefont{Stan, Romaniello,
  Rigamonti, Reining, and Berger}}]{stan}
\bibinfo{author}{\bibfnamefont{A.}~\bibnamefont{Stan}},
  \bibinfo{author}{\bibfnamefont{P.}~\bibnamefont{Romaniello}},
  \bibinfo{author}{\bibfnamefont{S.}~\bibnamefont{Rigamonti}},
  \bibinfo{author}{\bibfnamefont{L.}~\bibnamefont{Reining}}, \bibnamefont{and}
  \bibinfo{author}{\bibfnamefont{J.~A.} \bibnamefont{Berger}},
  \bibinfo{journal}{New Journal of Physics} \textbf{\bibinfo{volume}{17}},
  \bibinfo{pages}{093045} (\bibinfo{year}{2015}),
  \urlprefix\url{http://stacks.iop.org/1367-2630/17/i=9/a=093045}.

\bibitem[{\citenamefont{Berger et~al.}(2014)\citenamefont{Berger, Romaniello,
  Tandetzky, Mendoza, Brouder, and Reining}}]{berger2014}
\bibinfo{author}{\bibfnamefont{J.~A.} \bibnamefont{Berger}},
  \bibinfo{author}{\bibfnamefont{P.}~\bibnamefont{Romaniello}},
  \bibinfo{author}{\bibfnamefont{F.}~\bibnamefont{Tandetzky}},
  \bibinfo{author}{\bibfnamefont{B.~S.} \bibnamefont{Mendoza}},
  \bibinfo{author}{\bibfnamefont{C.}~\bibnamefont{Brouder}}, \bibnamefont{and}
  \bibinfo{author}{\bibfnamefont{L.}~\bibnamefont{Reining}},
  \bibinfo{journal}{New Journal of Physics} \textbf{\bibinfo{volume}{16}},
  \bibinfo{pages}{113025} (\bibinfo{year}{2014}),
  \urlprefix\url{http://stacks.iop.org/1367-2630/16/i=11/a=113025}.

\bibitem[{\citenamefont{Hedin}(1965{\natexlab{b}})}]{hedinPR65}
\bibinfo{author}{\bibfnamefont{L.}~\bibnamefont{Hedin}},
  \bibinfo{journal}{Phys. Rev.} \textbf{\bibinfo{volume}{139}},
  \bibinfo{pages}{A796} (\bibinfo{year}{1965}{\natexlab{b}}).

\bibitem[{\citenamefont{Caruso et~al.}(2012)\citenamefont{Caruso, Rinke, Ren,
  Scheffler, and Rubio}}]{caruso2012}
\bibinfo{author}{\bibfnamefont{F.}~\bibnamefont{Caruso}},
  \bibinfo{author}{\bibfnamefont{P.}~\bibnamefont{Rinke}},
  \bibinfo{author}{\bibfnamefont{X.}~\bibnamefont{Ren}},
  \bibinfo{author}{\bibfnamefont{M.}~\bibnamefont{Scheffler}},
  \bibnamefont{and} \bibinfo{author}{\bibfnamefont{A.}~\bibnamefont{Rubio}},
  \bibinfo{journal}{Phys. Rev. B} \textbf{\bibinfo{volume}{86}},
  \bibinfo{pages}{081102} (\bibinfo{year}{2012}),
  \urlprefix\url{https://link.aps.org/doi/10.1103/PhysRevB.86.081102}.

\bibitem[{\citenamefont{Holm and von Barth}(1998)}]{holm1998}
\bibinfo{author}{\bibfnamefont{B.}~\bibnamefont{Holm}} \bibnamefont{and}
  \bibinfo{author}{\bibfnamefont{U.}~\bibnamefont{von Barth}},
  \bibinfo{journal}{Phys. Rev. B} \textbf{\bibinfo{volume}{57}},
  \bibinfo{pages}{2108} (\bibinfo{year}{1998}),
  \urlprefix\url{https://link.aps.org/doi/10.1103/PhysRevB.57.2108}.

\bibitem[{\citenamefont{Guzzo et~al.}(2011)\citenamefont{Guzzo, Lani, Sottile,
  Romaniello, Gatti, Kas, Rehr, Silly, Sirotti, and Reining}}]{guzzo2011}
\bibinfo{author}{\bibfnamefont{M.}~\bibnamefont{Guzzo}},
  \bibinfo{author}{\bibfnamefont{G.}~\bibnamefont{Lani}},
  \bibinfo{author}{\bibfnamefont{F.}~\bibnamefont{Sottile}},
  \bibinfo{author}{\bibfnamefont{P.}~\bibnamefont{Romaniello}},
  \bibinfo{author}{\bibfnamefont{M.}~\bibnamefont{Gatti}},
  \bibinfo{author}{\bibfnamefont{J.~J.} \bibnamefont{Kas}},
  \bibinfo{author}{\bibfnamefont{J.~J.} \bibnamefont{Rehr}},
  \bibinfo{author}{\bibfnamefont{M.~G.} \bibnamefont{Silly}},
  \bibinfo{author}{\bibfnamefont{F.}~\bibnamefont{Sirotti}}, \bibnamefont{and}
  \bibinfo{author}{\bibfnamefont{L.}~\bibnamefont{Reining}},
  \bibinfo{journal}{Phys. Rev. Lett.} \textbf{\bibinfo{volume}{107}},
  \bibinfo{pages}{166401} (\bibinfo{year}{2011}),
  \urlprefix\url{https://link.aps.org/doi/10.1103/PhysRevLett.107.166401}.

\bibitem[{\citenamefont{{Sky Zhou} et~al.}(2017)\citenamefont{{Sky Zhou},
  {Gatti}, {Kas}, {Rehr}, and {Reining}}}]{sky2017}
\bibinfo{author}{\bibfnamefont{J.}~\bibnamefont{{Sky Zhou}}},
  \bibinfo{author}{\bibfnamefont{M.}~\bibnamefont{{Gatti}}},
  \bibinfo{author}{\bibfnamefont{J.~J.} \bibnamefont{{Kas}}},
  \bibinfo{author}{\bibfnamefont{J.~J.} \bibnamefont{{Rehr}}},
  \bibnamefont{and}
  \bibinfo{author}{\bibfnamefont{L.}~\bibnamefont{{Reining}}},
  \bibinfo{journal}{ArXiv e-prints}  (\bibinfo{year}{2017}),
  \eprint{1708.04313}.

\end{thebibliography}

\end{document}